\documentclass[]{aa} 

\usepackage{natbib}
\bibpunct{(}{)}{;}{a}{}{,} % to follow the A&A style
\usepackage{epsfig}

\newcommand{\beq}{\begin{equation}} 
\newcommand{\eeq}{\end{equation}} 
\newcommand{\beqa}{\begin{eqnarray}} 
\newcommand{\eeqa}{\end{eqnarray}} 
\newcommand{\bea}{\begin{array}} 
\newcommand{\ea}{\end{array}} 

\newcommand{\dd}{{\rm d}}

\newcommand{\lag}{\langle} 
\newcommand{\rag}{\rangle} 

\newcommand{\ii}{{\rm i}}

\newcommand{\Om}{\Omega_{\rm m}} 
\newcommand{\OL}{\Omega_{\Lambda}} 
\newcommand{\rhob}{\overline{\rho}}

\newcommand{\deltas}{\delta_{*}}
 \newcommand{\deltaLs}{\delta_{L*}}

\newcommand{\nb}{\overline{n}}

\newcommand{\vk}{{\bf k}}
\newcommand{\vPsi}{{\bf \Psi}}
\newcommand{\vq}{{\bf q}}
\newcommand{\vs}{{\bf s}}

\newcommand{\vx}{{\bf x}}

\newcommand{\tdelta}{{\tilde{\delta}}}
\newcommand{\tW}{\tilde{W}}

\newcommand{\cF}{{\cal F}}
\newcommand{\cP}{{\cal P}}

\begin{document}

\title{Large-scale bias of dark matter halos}    
\author{P. Valageas}   
\institute{Institut de Physique Th\'eorique, CEA Saclay, 91191 Gif-sur-Yvette, 
France}  
\date{Received / Accepted } 
 
\abstract
{}
{We build a simple analytical model for the bias of dark matter halos that applies
to objects defined by an arbitrary density threshold, $200\leq\deltas\leq 1600$,
and that provides accurate predictions from low-mass to high-mass halos.}
{We point out that it is possible to build simple and efficient models, with
no free parameter for the halo bias, by using integral constraints that govern
the behavior of low-mass and typical halos, whereas the properties of rare
massive halos are derived through explicit asymptotic approaches. We also
describe how to take into account the impact of halo motions on their bias,
using their linear displacement field.}
{We obtain a good agreement with numerical simulations for the halo mass
functions and large-scale bias at redshifts $0\leq z \leq 2.5$, for halos defined
by a nonlinear density threshold $200\leq\deltas\leq 1600$. We also evaluate the
impact on the halo bias of two common approximations, i) neglecting halo
motions, and ii) linearizing the halo two-point correlation.}
{}


\maketitle

\section{Introduction} 
\label{Introduction}

A fundamental test of cosmological models is provided by the distribution of
nonlinear virialized objects, such as galaxies or clusters of galaxies.
First, this allows to check that nonlinear objects form through the amplification
by gravitational instability of small (nearly Gaussian) primordial density
fluctuations \citep{Peebles1980}. Second, quantitative comparisons between
theoretical predictions and observations allow to derive constraints on
cosmological parameters \citep{Zheng2007}.
Indeed, observations show that galaxies and clusters do not follow a Poisson
distribution but show significant large-scale correlations (e.g.,
\citet{McCracken2008,Padilla2004}). In particular, at large scales their two-point
correlation function is roughly proportional to the underlying matter correlation,
with a multiplicative factor $b^2$, called the bias, that is greater for more
massive objects.

Following the spirit of the Press-Schechter picture \citep{Press1974}, where
nonlinear virialized objects are identified with high-density fluctuations in the
initial (linear) density field, \citet{Kaiser1984} showed how such a behavior
for the halo correlation naturally arises. Similar results are obtained for the
clustering of density peaks instead of overdensities above a given
threshold \citep{Bardeen1986}. Indeed, distant high-density regions
are correlated through their common longwavelength modes of the linear
density field, and the effect is greater for more massive and extreme objects.
A somewhat simpler derivation can be obtained through the peak-background
split argument \citep{Cole1989,Bond1991}. This rests on the same physics,
which is that within a large region characterized by a positive mean density
contrast local density peaks beyond a given density threshold are more frequent
than in the mean, which yields a positive correlation between rare massive halos
and the larger-scale matter density field.

Along these lines, a simple analytical model that compares reasonably well
with numerical simulations was presented by \citet{MW1996}, while a better
agreement was obtained by \citet{Sheth1999}, at the cost of adding a few
parameters fitted to the simulations (through the halo mass function).
More complex models, that also apply to redshift space and take into account
subleading scale-dependent terms can be found for instance in
\citet{Desjacques2008,Desjacques2010}.
On the other hand, various fitting formulas have been proposed from measures
in numerical simulations
\citep{Hamana2001,Pillepich2010,Tinker2010,Manera2010}.
In particular, \citet{Tinker2010} have recently studied the dependence of
the halo bias on the density contrast $\deltas$ used to define the halos.
In the paper we present a simple analytical model for the halo bias that extends
a previous work \citep{Valageas2009d} to arbitrary density thresholds in the
range $200\leq \deltas\leq 1600$. This is of practical interest as the
definition of halos can vary among authors, and it can happen that observations
only extend to smaller radii than the usual virial radius, which corresponds to
objects defined by higher density thresholds.
In addition, we also improve on \citet{Valageas2009d} by simplifying somewhat
the model, especially the treatment of halo motions, while obtaining a better
accuracy for low-mass halos.
This should allow an easier generalization to more complex cases, such as
non-Gaussian initial conditions.
This simple model also allows us to evaluate the inaccuracies implied by two 
approximations that are frequently used, that is, neglecting halo motions and
linearizing the halo correlation function over the matter correlation function.

In Sect.~\ref{Bias} we describe our model for the bias of halos defined by a
nonlinear density contrast $\deltas=200$. We explain how the standard model
of \citet{Kaiser1984} for the correlation of rare massive halos is modified
once we take into account the motion of halos. Then, we point out that
normalization conditions can be very useful to constrain the bias of intermediate
and low mass halos, which allows to build a simple and efficient model.
Next, we show that we obtain a good agreement with numerical simulations.
In Sect.~\ref{Extension} we extend this model for the bias, as well as for the halo
mass function, to the case of halos defined by a larger nonlinear density contrast,
and we compare our results with numerical simulations for
$200\leq \deltas\leq 1600$. 
Finally, we estimate the importance of halo motions and of the nonlinearity
of the bias in Sects.~\ref{Impact-of-halo-motions} and
\ref{Impact-of-exponential-nonlinearity}, and we conclude in
Sect.~\ref{Conclusion}.

\section{Bias of dark matter halos}
\label{Bias}

Following \citet{Kaiser1984} we estimate the two-point correlation function of dark
matter halos, whence their bias, from the probability to reach a linear density
threshold $\deltaLs$ (associated with the formation of virialized halos of nonlinear
density contrast $\deltas$). As in \citet{Valageas2009d}, we improve this model
by taking into account the motion of halos, associated with the change from
Lagrangian to Eulerian space (i.e., from the linear density field to the actual
nonlinear density field). However, we simplify the prescription used in
\citet{Valageas2009d} as we no longer use a spherical collapse model to estimate
these displacements but use the initial momenta of both halos (i.e., the linear
displacement field). In addition, we add a further ingredient to the study of
\citet{Valageas2009d}, as we explicitly enforce the normalization to unity of the
halo bias (when integrated over all halo masses).
We describe below this simple model.

\subsection{Lagrangian space}
\label{Lagrangian-space}

We first consider the two-point correlation function of halos in Lagrangian space,
that is within the linear density field $\delta_L(\vq)$, where we identify future
halos of Eulerian radius $r$ and nonlinear density contrast $\deltas$
(with typically $\deltas\sim 200$) with spherical regions of Lagrangian radius $q$
and linear density contrast $\deltaLs$.
Thanks to the conservation of mass, the Lagrangian and Eulerian properties of halos
of mass $M$ are related by
\beq
M= \frac{4\pi}{3} \rhob \, q^3, \hspace{0.45cm} q^3 = (1+\deltas) \, r^3 
\hspace{0.45cm} \mbox{and} \hspace{0.45cm} \deltas=\cF(\deltaLs) ,
\label{qr-F}
\eeq
where $\rhob$ is the mean matter density.
As pointed out in \citet{Valageas2009d}, the function $\cF$ describes the
spherical collapse dynamics, so that in a $\Lambda$CDM cosmology with
$\Om=0.27$, $\OL=0.73$, we have $\deltaLs \simeq 1.59$ at $z=0$ for
$\deltas=200$, instead of the usual value $\delta_c\simeq 1.675$ associated
with full collapse to a point, that is with $\deltas=\infty$.

Defining halos by the linear threshold $\deltaLs=\cF^{-1}(\deltas)$, rather
than by the full collapse value $\delta_c$, is closer to observational and numerical
procedures, since (in the best cases) one defines ``halos'' in galaxy or cluster
surveys, and in numerical simulations, by the radius and mass of overdensities
within a given nonlinear density threshold $\deltas$. Moreover, this gives the
freedom to chose different thresholds, such as $\deltas=200$ or $100$
\citep{Valageas2009d}. As discussed in Sect.3 in \citet{Valageas2009d},
for massive ($M\sim 10^{15} h^{-1}M_{\odot}$) and rare halos, the choice
$\deltas=200$ also roughly corresponds to the separation between outer shells,
dominated by radial accretion, and inner shells, with a significant transverse
velocity dispersion, that have experienced shell crossing. 
The case of typical halos (i.e. below the knee of the halo mass function) is more
intricate as they do not show such a clear separation. This can also be seen in
numerical simulations, such as Fig.3 in \citet{Cuesta2008}. At high redshift,
where nonlinear objects have a smaller mass, this ``virialization'' radius shifts to
higher density contrasts for $\Lambda$CDM cosmologies, because of the change of
slope of the linear matter power spectrum with scale (e.g., $\deltas \sim 500$ for
$M\sim 10^{11} h^{-1}M_{\odot}$, as seen in Fig.5 in \citet{Valageas2009d}).

Then, defining the halo correlation as the fractional excess of halo pairs
\citep{Kaiser1984,Peebles1980}, we write in Lagrangian space
\beqa
\nb_L(M_1,M_2;\vs_1,\vs_2) \dd M_1 \dd M_2 \dd \vs_1 \dd \vs_2 & = &
\nonumber \\
&& \hspace{-5.4cm} \nb_L(M_1) \nb_L(M_2) [ 1 + \xi_L(M_1,M_2;s) ] 
\dd M_1 \dd M_2 \dd \vs_1 \dd \vs_2 .
\label{xiL-def}
\eeqa
Here and in the following we use the letter $\vs$ for the position of halos
in Lagrangian space to avoid confusion with their Lagrangian radius $q$,
and we introduced the Lagrangian distance, $s=|\vs_2-\vs_1|$, between
both objects. We also note with a subscript $L$ the quantities associated
with the linear fields or the Lagrangian space.
Then, following \citet{Kaiser1984} and \citet{Valageas2009d}, we obtain the
fractional excess of halo pairs from the bivariate density distribution
$\cP_L(\delta_{L1},\delta_{L2})$ over the two spheres of radii $q_1$ and $q_2$,
\beqa
1 + \xi_L(M_1,M_2;s) & = & \frac{\cP_L(\delta_{L1},\delta_{L2})}
{\cP_L(\delta_{L1})\cP_L(\delta_{L2})} \\
& &  \hspace{-3cm} = \frac{\sigma_1\sigma_2}
{\sqrt{\sigma_1^2 \sigma_2^2\!-\!\sigma_{12}^4}}
\, \exp\left(\frac{\deltaLs^2 \sigma_{12}^2 (2\!-\!\sigma_{12}^2/\sigma_1^2
\!-\!\sigma_{12}^2/\sigma_2^2)}{2(\sigma_1^2\sigma_2^2-\sigma_{12}^4)}
\right) ,
\label{xiL-s}
\eeqa
where we considered a single population, that is halos defined by the same
density thresholds $\deltaLs$ and $\deltas=\cF(\deltaLs)$.
Here we assumed Gaussian initial conditions and we introduced the
cross-correlation of the smoothed linear density contrast at scales $q_1$
and $q_2$, at positions $\vs_1$ and $\vs_2$,
\beqa
\sigma^2_{q_1,q_2}(s) & = & \lag \delta_{Lq_1}(\vs_1)
\delta_{Lq_2}(\vs_1+\vs) \rag  \nonumber \\
& = & 4\pi\int_0^{\infty} \dd k \, k^2 \, P_L(k) \tW(kq_1) \tW(kq_2) 
\frac{\sin(ks)}{ks} ,
\label{sigq1q2}
\eeqa
where $\tW(kq)$ is the Fourier transform of the top-hat window of radius $q$,
\beq
\tW(kq) = \int_V \frac{\dd \vq}{V} \, e^{\ii\vk\cdot\vq} = 
3 \frac{\sin(kq)-kq\cos(kq)}{(kq)^3} ,
\label{Wdef}
\eeq
and $P_L(k)$ is the linear matter power spectrum, defined by
\beq
\delta_L(\vs) = \int\dd\vk \, e^{\ii\vk\cdot\vs} \, \tdelta_L(\vk) ,
\label{Fourierdef}
\eeq
\beq
\lag \tdelta_L(\vk_1)\tdelta_L(\vk_2) \rag = \delta_D(\vk_1+\vk_2) \, P_L(k_1) .
\label{PLdef}
\eeq
In particular, $\sigma_q=\sigma_{q,q}(0)$ is the usual rms linear density
contrast at scale $q$.
In Eq.(\ref{xiL-s}) we also used the short-hand notation
$\sigma_1^2=\sigma^2_{q_1,q_1}(0)$, $\sigma_2^2=\sigma^2_{q_2,q_2}(0)$,
and $\sigma_{12}^2=\sigma^2_{q_1,q_2}(s)$, for the covariances over the
two spherical Lagrangian regions.

Of course, the prescription (\ref{xiL-s}) only applies to the limit of rare
events (i.e., large mass $M$) and large separations, where halos are isolated 
and almost spherical, so that we can neglect tidal effects as well as 
violent processes such as mergings. We shall come back to this point in
Sect.~\ref{Normalization} below.
The Eq.(\ref{xiL-s}) expresses the fact that if we have a rare overdensity
at position $\vs_1$ the probability to have a second overdensity at a nearby
position $\vs_2$ is amplified, with respect to random locations, because
of common longwavelength modes.

\subsection{Eulerian space}
\label{Eulerian-space}

We must now convert the expression (\ref{xiL-s}) into a model for the halo
correlation in Eulerian space, that is within the actual nonlinear density
field. Indeed, even halos that have not been destroyed or strongly modified
by merging events have had time to move since the early times associated
with the linear density field (which formally corresponds to $z\rightarrow\infty$).
In particular, the particles they are made of have moved even before the formation
of the final halo. This leads to two effects with respect to the halo correlation:
i) the Eulerian-space distance $x$ between the two halos is not identical to the
Lagrangian-space distance $s$, ii) the halo number densities are modified because
the Lagrangian-Eulerian mapping modifies local volumes (i.e., $\dd\vx\neq\dd\vs$).
These effects were estimated in \citet{Valageas2009d} from a spherical collapse
model. Here we present a simpler method, which has the advantage of being more
flexible and more closely related to the estimate (\ref{xiL-s}), based on the
linear density field. Moreover, it avoids the need to use a spherical dynamics
approximation. As a further simplification, we also neglect the second
effect ii), due to the change of volume between Eulerian and Lagrangian
spaces, because it is a subdominant prefactor.

Thus, we simply estimate the displacement $\vPsi(M,\vs)$ of a halo of mass $M$
and Lagrangian position $\vs$ from the linear displacement field (see also
\citet{Desjacques2008} for a somewhat different implementation in the
case of density peaks).
This gives for the Eulerian position of its center of mass,
\beq
\vx(M,\vs) = \vs + \vPsi_L(M,\vs) ,
\label{x-Ms}
\eeq
with
\beq
\vPsi_L(M,\vs) = \ii \int \dd\vk \, e^{\ii\vk\cdot\vs} \, \frac{\vk}{k^2} \,
\tdelta_L(\vk) \tW(k q) .
\label{Psi-Ms}
\eeq
This coincides with the use of the Zeldovich approximation \citep{Zeldovich1970}
for the dynamics of the particles that make up the halo of mass $M$.
However, since the halo scale $q$ introduces a natural cutoff at high $k$
(through the factor $\tW(k q)$), which for large masses and rare events is
within the linear regime, the approximation (\ref{x-Ms}) should fare much better
than what could be expected from an analysis of the Zeldovich dynamics 
for individual particle trajectories. For instance, the ``truncated Zeldovich
approximation'', where one suppresses the initial power at high $k$, provides
a good description of large-scale clustering and yields a significant improvement
over the unsmoothed Zeldovich approximation \citep{Coles1993,Melott1994a}.
Physically, this means that small-scale nonlinear processes such as virialization
within bound objects conserve momentum and do not affect the large-scale
properties of the system.

Then, for Gaussian initial conditions we can obtain the trivariate probability
distribution $\cP_L(\delta_{L1},\delta_{L2},\Psi_{L12\parallel})$, where
$\Psi_{12\parallel}$ is the linear relative displacement of the halos along their
Lagrangian separation vector,
\beq
\Psi_{L12\parallel} = \frac{[\vPsi_L(M_2,\vs_2)-\vPsi_L(M_1,\vs_1)]\cdot
(\vs_2-\vs_1)}{|\vs_2-\vs_1|} .
\label{Psi12-def}
\eeq
In particular, the mean conditional relative displacement, at fixed linear
density contrasts $\delta_{L1}$ and $\delta_{L2}$ within the two spheres
of radii $q_1$ and $q_2$, reads as
\beq
\lag \Psi_{L12\parallel} \rag_{\delta_{L1},\delta_{L2}} = 
- \sigma_{\delta_L\Psi_{L\parallel}}^2 
\frac{\delta_{L1}(\sigma^2_2\!-\!\sigma^2_{12})
+\delta_{L2}(\sigma^2_1\!-\!\sigma^2_{12})}{\sigma_1^2\sigma_2^2-\sigma_{12}^4} ,
\label{Psi12-dL}
\eeq
where we introduced the covariance
\beq
\sigma_{\delta_L\Psi_{L\parallel}}^2 = - \lag \delta_{L1} \Psi_{L12\parallel} \rag
= - \lag \delta_{L2} \Psi_{L12\parallel} \rag ,
\label{sigma-par}
\eeq
which writes as
\beq
\sigma_{\delta_L\Psi_{L\parallel}}^2 = \frac{s}{3} \, \sigma^2_{q_1,q_2,s} 
\label{sigma-3}
\eeq
with
\beq
\sigma^2_{q_1,q_2,s} = 4\pi\int_0^{\infty} \dd k \, k^2 \, P_L(k)
\tW(kq_1) \tW(kq_2) \tW(ks) ,
\label{sigq1q2s}
\eeq
where again $s=|\vs_2-\vs_1|$.
We introduced a minus sign in the definition (\ref{sigma-par}) to avoid
a minus sign in Eqs.(\ref{sigma-3})-(\ref{sigq1q2s}). Although
$\sigma^2_{q_1,q_2,s}$ is not guaranteed to be always positive, it is usually
positive, which expresses through (\ref{sigma-par}) that in the mean two rare
overdensities tend to move closer, as could be expected through their mutual
gravitational attraction.
By symmetry, the mean conditional relative displacement along the
orthogonal directions vanishes,
$\lag \Psi_{L12\perp} \rag_{\delta_{L1},\delta_{L2}} = 0$.
Then, at lowest order the mean Eulerian-space distance $x$ between both halos
reads as
\beq
\lag x(M_1,M_2;s) \rag = s- \frac{s}{3} \, \deltaLs \, \sigma^2_{q_1,q_2,s}
\frac{\sigma^2_1+\sigma^2_2-2\sigma^2_{12}}
{\sigma_1^2\sigma_2^2-\sigma_{12}^4} ,
\label{x-s}
\eeq
where we considered a single population defined by the same density
threshold $\deltaLs$.
Neglecting the width of the distribution $\cP_L(\Psi_{L12\parallel})$, whence
of $\cP_L(x)$, and in the limit of large separation, $s\rightarrow \infty$, where
$\sigma^2_{q_1,q_2,s}\rightarrow 0$, we can invert Eq.(\ref{x-s}) as
\beq
s(M_1,M_2;x) \simeq x \left[ 1 + \frac{\deltaLs}{3} 
\frac{\sigma^2_{q_1,q_2,x}(\sigma^2_1+\sigma^2_2-2\sigma^2_{12})}
{\sigma_1^2\sigma_2^2-\sigma_{12}^4} \right] ,
\label{s-x}
\eeq
where $\sigma^2_{12}$ is also evaluated at distance $x$.

It is interesting to note that the expression (\ref{s-x})
is very close to the one obtained in \citet{Valageas2009d} from a spherical
collapse model. The main difference is that this new result involves the
quantity $\sigma^2_{q_1,q_2,s}$ defined in Eq.(\ref{sigq1q2s}), which contains
the three windows $\tW(kq_1), \tW(kq_2)$, and $\tW(ks)$, whereas the
former result only involved quantities such as $\sigma^2_{q,s}$ and
$\sigma^2_{q,0}(s)$ that only contain two windows.
The reason is that the model used in \citet{Valageas2009d} treated each
halo as a test particle within the spherical gravitational field built by the other
halo, whereas in the derivation of Eq.(\ref{s-x}) above
we did not need to make this approximation and we simultaneously take into
account the effect of both halos on the initial gravitational field.
Nevertheless, it is reassuring that both models give similar results, as could
be expected from qualitative arguments (for rare and well-separated halos).

Equation (\ref{s-x}) gives the ``typical'' Lagrangian-space distance $s$ between
two halos that are observed in the nonlinear density field at the Eulerian
distance $x$. This takes care of the first effect i) associated with halo motions.
As noticed above, the second effect ii) of halo displacements is to change
volumes so that $\dd\vx\neq\dd\vs$. Then, by conservation of the number
of pairs, the Lagrangian-space and Eulerian-space correlations at distances
$\vs$ and $\vx$ are related by
\beq
[1+\xi(M_1,M_2;x)] \dd\vx = [1+\xi_L(M_1,M_2;s)] \dd\vs ,
\eeq
and we could use again Eqs.(\ref{x-Ms}) and (\ref{s-x}) to estimate the factor
$|\dd\vs/\dd\vx|$. However, for rare massive halos, of Lagrangian radius $q$,
this is a subdominant effect, since from Eq.(\ref{s-x}) we have
$|\Delta s / \Delta x| \sim \deltaLs \sigma^2_{q,q,s}/\sigma_q^2$,
whereas the argument of the exponential (\ref{xiL-s}) behaves as
$\deltaLs^2 \sigma^2_{12}/\sigma_q^4$. Thus, the latter is larger than
the former by a factor $\deltaLs/\sigma_q^2$, which is much greater than unity
(and this is further amplified for very rare halos by the exponential).
Therefore, to avoid introducing unnecessary approximations and to make
the model as simple as possible, we neglect this volume effect and we only keep
the exponential behavior of the two-point correlation of rare halos. This
gives for equal-mass halos, with $M_1=M_2=M$ and $q_1=q_2=q$,
\beq
1+\xi(M,x) \simeq e^{\deltaLs^2 \sigma^2_{q,q}(s)/
[\sigma^2_q(\sigma^2_q+\sigma^2_{q,q}(s))]}  ,
\label{xi-M}
\eeq
whereas the Lagrangian distance reads from Eq.(\ref {s-x}) as
\beq
s(M,x) = x \left( 1+\frac{2 \, \deltaLs \, \sigma^2_{q,q,x}}
{3\,(\sigma^2_q+\sigma^2_{q,q}(s))} \right) .
\label{s-x-M}
\eeq
Then, since at large distance the dark matter two-point correlation function
is within the linear regime and reads as $\xi(x) \simeq \sigma^2_{0,0}(x)$,
we obtain for the bias of halos of mass $M$,
\beqa
b^2_{\rm r.e.}(M,x) & = & \frac{\xi(M,x)}{\xi(x)} \\
& = & \frac{1}{\sigma^2_{0,0}(x)} \left[ e^{\deltaLs^2 \sigma^2_{q,q}(s)/
[\sigma^2_q(\sigma^2_q+\sigma^2_{q,q}(s))]} -1 \right] .
\label{b2-def}
\eeqa
Here the subscript ``r.e.'' stands for the ``rare-event'' limit to recall that the
results obtained so far have been derived for very massive and rare halos, and
do not apply to small objects.

\subsection{Normalization}
\label{Normalization}

It is very difficult to derive an approximate model for the bias, and even
the mass function, of small halos. Indeed, such objects associated with
typical density fluctuations have been strongly distorted by tidal effects
and have often undergone various merging events. Then, it is not possible
to identify in a simple manner the precursors of these halos in the initial
linear density field. In this article we propose the following simple prescription
to bypass this problem: we add to the asymptotic result (\ref{b2-def})
a constant term with respect to the halo mass, $b_0(x)$,  which we obtain from
the normalization of the bias. More precisely, making the approximation that
the bias factorizes (which however is not exact, even in the large-mass regime,
see Eq.(\ref{b2-def}) and Fig.~12 in \citet{Valageas2009d}), as
\beq
b^2(M_1,M_2;x) \simeq b(M_1,x) \, b(M_2,x) ,
\label{b-fact}
\eeq
we model the bias $b(M,x)$ as
\beq
b(M,x) =   b_{\rm r.e.}(M,x) + b_0(x) .
\label{bM-def}
\eeq
Here $b_{\rm r.e.}(M,x)$ is given by Eq.(\ref{b2-def}) (taking the square-root of
this positive quantity), while $b_0(x)$ is set by the normalization (at fixed
distance $x$)
\beq
\int_0^{\infty} b(M,x) \, f(\nu) \, \frac{\dd \nu}{\nu} = 1 ,
\label{b-fnu}
\eeq
whence
\beq
b_0(x) = 1 - \int_0^{\infty} b_{\rm r.e.}(M,x) \, f(\nu) \, \frac{\dd \nu}{\nu} ,
\label{b0-def}
\eeq
where we used
\beq
\int_0^{\infty} f(\nu) \, \frac{\dd \nu}{\nu} = 1 .
\label{fnu-norm}
\eeq
Here we introduced the scaled mass function of dark matter halos, that is,
we write the halo mass function as
\beq
n(M) \dd M = \frac{\rhob}{M} \, f(\nu) \, \frac{\dd\nu}{\nu} \;\;\;
\mbox{with} \;\;\; \nu=\frac{\deltaLs}{\sigma(M)} ,
\label{fnu-def}
\eeq
where $\deltaLs$ is the linear density threshold that defines these halos,
given by Eq.(\ref{qr-F}), and $\sigma(M)=\sigma_q$.
The well-know constraints (\ref{fnu-norm}) and (\ref{b-fnu}) follow from the
requirement that
all the mass is accounted for by the mass function of halos (whatever the
choice of $\deltaLs$), which leads to Eq.(\ref{fnu-norm}),
so that the density field on large scales (beyond the
size of these halos) can be written in terms of the halo mass function,
as in the usual halo model \citep{Cooray2002} (the so-called ``two-halo term'').
Then, requiring that one recovers the dark matter two-point correlation
function leads to the constraint (\ref{b-fnu}), provided the bias 
can be factorized as (\ref{b-fact}) (otherwise the constraint involves
a bidimensional integral over $b^2(M_1,M_2)$).

We do not claim here that the halo bias has a nonzero asymptote at low
mass. In fact, the asymptotic behavior of the low-mass tail of the bias (and
of the mass function itself) is largely unknown, but numerical simulations
show that the dependence on mass flattens below $\nu \sim 1$
and seems roughly constant down to $\nu \sim 0.3$. Then, the prescription
(\ref{bM-def}) is intended to describe both this behavior
(since from expression (\ref{b2-def}) we can see that
$b_{\rm r.e.}(M) \rightarrow 0$ for $M\rightarrow 0$, or more precisely when
$\sigma_q \rightarrow \infty$) and the constraint (\ref{b-fnu}).
Indeed, if the asymptotic behavior (\ref{b2-def}) provides a good description
for $\nu > 1$ we can expect the constraint (\ref{b-fnu}) to provide a good
estimate for the (almost constant) value of $b(M)$ in the regime $\nu < 1$.
Moreover, it is always useful to make sure normalization constraints such
as (\ref{b-fnu}) are satisfied by the models. Indeed, this ensures that
the models are self-consistent and that absurd results will not be produced
by the violation of basic internal constraints.

Another advantage of this simple prescription is that our model (\ref{bM-def})
of the halo bias has {\it no} specific free parameter, apart from those
already contained in the halo mass function (especially its low-mass tail).
This is why we prefer to keep a simple constant term $b_0$, instead
of introducing for instance higher-order polynomials (over $\nu$ or $M$)
that would require some fitting over numerical simulations.
This provides a greater flexibility to the model, which can be used
for a variety of cosmologies.

\begin{figure*}[htb]
\begin{center}
\epsfxsize=6.05 cm \epsfysize=6 cm {\epsfbox{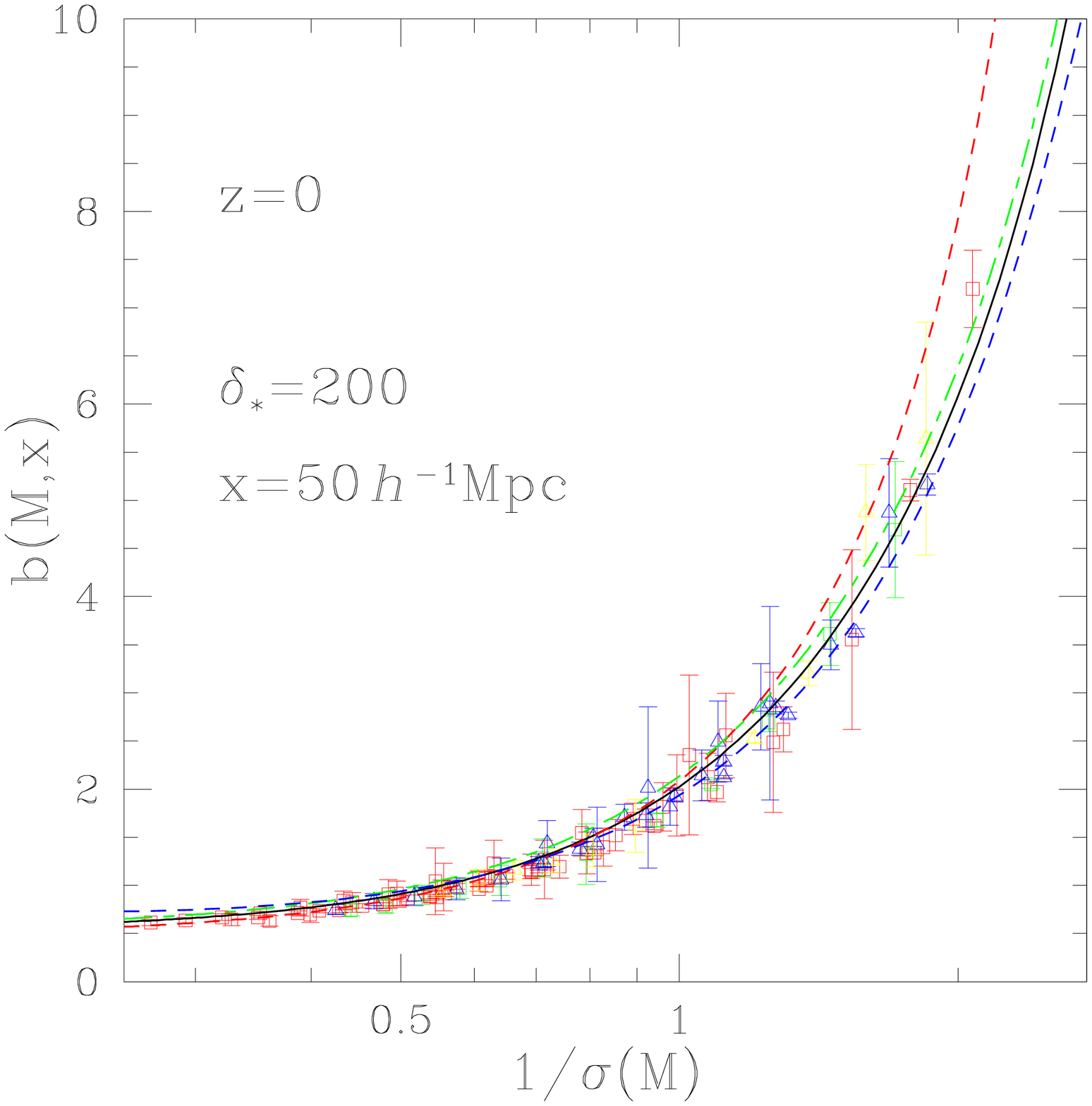}}
\epsfxsize=6.05 cm \epsfysize=6 cm {\epsfbox{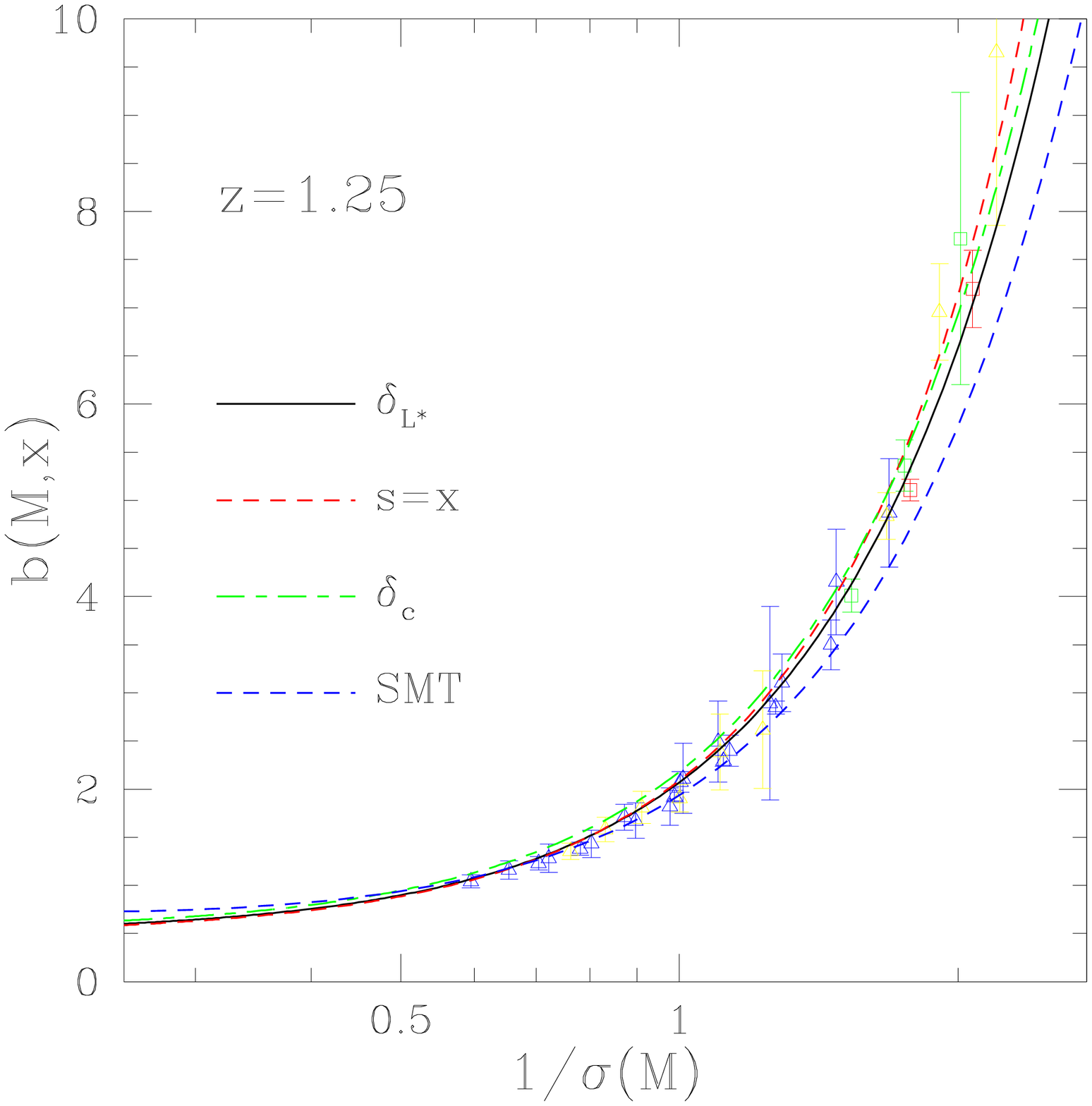}}
\epsfxsize=6.05 cm \epsfysize=6 cm {\epsfbox{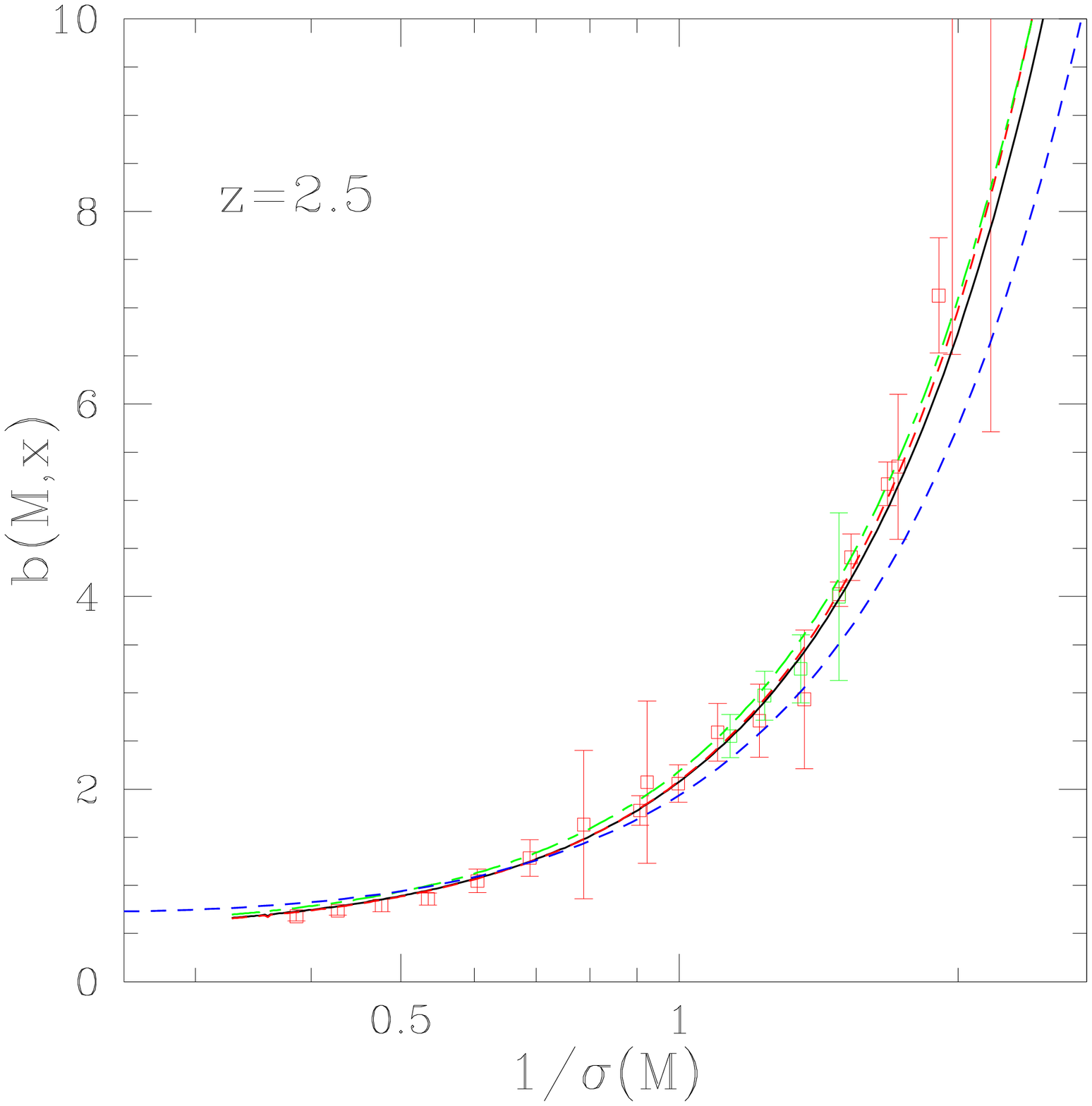}}
\end{center}
\caption{The halo bias $b(M,x)$ as a function of $\sigma(M)$, at distance
$x=50 h^{-1}$Mpc, for halos defined by the nonlinear density contrast
$\deltas=200$. We plot our results at redshifts $z=0$ (left panel),
$z=1.25$ (middle panel), and $z=2.5$ (right panel).
The black solid line, ``$\deltaLs$'', is the theoretical prediction (\ref{bM-def}),
the red dashed line, ``$s=x$'', corresponds to the approximation where
halo motions are neglected, and the green dot-dashed line, ``$\delta_c$'',
corresponds to the use of $\delta_c=1.686$ instead of $\deltaLs$.
The blue dashed line, ``SMT'', is the fit from \citet{SMT2001}, and the points are
the results from numerical simulations of \citet{Tinker2010}.}
\label{figbiasM_D200}
\end{figure*}

Finally, for numerical computations we use the mass function given in
\citet{Valageas2009d},
\beq
f(\nu) =  0.502 \left[ (0.6 \,\nu)^{2.5} + (0.62 \, \nu)^{0.5} \right] \, e^{-\nu^2/2} ,
\label{fnu-fit}
\eeq
which has been shown to agree with numerical simulations (for halos
defined by $\deltas=200$).
The exponential falloff, $e^{-\nu^2/2}$, where $\nu$ is defined by
Eqs.(\ref{fnu-def}) and (\ref{qr-F}), is consistent with the exponential
term of Eq.(\ref{xiL-s}). Indeed, both the 1-point distribution (i.e. the mass
function) and the 2-point distribution (i.e. the halo correlation or bias)
are obtained in the large-mass limit from spherical overdensities
$\deltaLs$ in the linear density field, with $\deltaLs=\cF^{-1}(\deltas)$
and $\deltas=200$.
Note also that the normalization of the mass function (\ref{fnu-fit}) is not
a free parameter since it is set by the constraint (\ref{fnu-norm}).
This is why we prefer to use the mass function (\ref{fnu-fit}), so that the
model is fully self-consistent and large-mass tails do not involve
free parameters.

In particular, following \citet{Cole1989}, applying the peak-background split
argument to Eq.(\ref{fnu-fit}) gives $b \sim \deltaLs/\sigma_q^2$, in the
rare-event and large-distance limits. This agrees with the asymptotic
behavior of Eq.(\ref{b2-def}), except that Eq.(\ref{b2-def}) also yields the prefactor
$\sigma_{q,q}(s)/\sigma_{0,0}(x)$, which is different from unity.
This expresses the facts that i) contributions to the correlation of objects of size
$q$ are damped for high wavenumbers, $k \gg 1/q$, and ii) halo motions are
nonzero and correlated, $s\neq x$.
Of course, these two effects are neglected by the peak-background split
argument.

\subsection{Comparison with numerical simulations}
\label{Comparison}

We now compare the model defined by Eqs.(\ref{s-x-M}), (\ref{b2-def}),
(\ref{bM-def}), and (\ref{b0-def}), with results from numerical simulations.
We first consider in Fig.~\ref{figbiasM_D200} the dependence on
halo mass $M$ of the large-scale bias, for halos defined
by the nonlinear density contrast $\deltas=200$ (whence
$\deltaLs\simeq 1.59$) at distance $x=50 h^{-1}$Mpc.
We show our results at redshifts $z=0, 1.25$, and $2.5$.
We can see that we obtain a good agreement with the numerical simulations
of \citet{Tinker2010}. This was expected at high masses, where the arguments
based on the clustering of rare overdensities in the linear Gaussian density
field apply (as introduced by \citet{Kaiser1984} and implemented in a
slightly modified variant here).
An improvement over previous
models of this kind \citep{Kaiser1984,Valageas2009d} is the good agreement
at low mass, which is obtained through the constant term $b_0$ in
Eq.(\ref{bM-def}), associated with the normalization of the halo bias
through Eq.(\ref{b0-def}). Thus, it appears that this constraint is sufficient
to obtain a good description of the halo bias over the whole range
$\sigma(M)<10$. 
Indeed, as explained in Sect.~\ref{Normalization}, since the model is reasonably
successful at large mass, $\sigma<1$, the integral constraint (\ref{b-fnu})
ensures that the value of the bias over the range $1<\sigma<10$,
which contains the other half of the total matter content, has the correct
magnitude, although the detailed shape has no reason to be exact
a priori. In particular, at low masses, $\sigma>10$, which represent a small
fraction of the total matter density field and do not significantly contribute
to the integral (\ref{b-fnu}), the normalization (\ref{b-fnu}) becomes largely
irrelevant (since large changes of $b(M)$ would not significantly change
the overall normalization) so that there is no reason a priori to trust our
model. Thus, it may happen that at very low masses the bias goes to zero,
for instance as a power law over $M$. However, this is beyond the reach
of current numerical simulations and it is not a serious practical problem,
for the same reason that this only concerns a small fraction of the matter
content and of the halo population.

As is well known, the dependence on redshift, at fixed $\sigma(M)$, is quite
weak over the range $0\leq z \leq 2.5$. There appears to be a slight
growth of the bias at larger redshift, in the regime of rare halos ($\sigma(M)<0.5$).
This is most clearly seen by the comparison with the dashed line
associated with the fit from \citet{SMT2001}, which is independent of $z$
and is identical in the three panels. This small growth is also reproduced
by our model (\ref{bM-def}), as shown by the solid line. This confirms
the validity of this approach, and more generally of such models that follow
\citet{Kaiser1984}. However, for practical purposes it is probably sufficient
to neglect the dependence on redshift in this range.

It is interesting to note that the bias obtained from Eq.(8) of \citet{SMT2001}
behaves at large masses as $b \sim a \delta_c/\sigma_q^2$, with a parameter 
$a\simeq 0.707$ obtained from fits to the halo mass function
(through the peak-background split argument), whereas we have
noticed in Sect.~\ref{Normalization} that Eq.(\ref{b2-def}) yields
$b \sim (\delta_{L*}/\sigma_q^2) (\sigma_{q,q}(s)/\sigma_{0,0}(x))$.
In order to explain why both predictions are rather close
(especially at $z=0$ in Fig.~\ref{figbiasM_D200}) despite these different
forms, and without introducing such a parameter $a$ in Eq.(\ref{b2-def}),
we have also plotted the curves obtained by setting $s=x$ (i.e. neglecting halo
motions) or $\deltaLs=\delta_c$, with the usual value $\delta_c=1.686$.
We can see that the change from $\delta_c$ to $\deltaLs$ makes almost
no difference for the halo bias (for $\deltas=200$), while taking into account
halo motions leads to a significant reduction at $z=0$ for massive halos.
Indeed, since rare halos tend to move closer, neglecting halo motions
underestimates their initial distance and overestimates their correlation.
This effect is greater at lower redshift, associated with more massive
objects, and leads to a bias that happens to be very close to the prediction from
\citet{SMT2001} at $z=0$.

Thus, taking into account halo motions has the same effect (i.e. reducing the bias
at low $z$) as the parameter $a$ of \citet{SMT2001}. Within the peak-background
split approach, the latter is not a new free parameter for the bias, since it is set by
the halo mass function itself. However, within our approach, we do not need to
introduce such a parameter, neither for the halo mass function (\ref{fnu-fit}) nor for
the bias (\ref{b2-def}).
We can note that the very weak dependence on redshift of the halo bias, in the
range $0\leq z \leq 2.5$, is somewhat misleading, since within this model it arises
from the partial cancellation between two opposite trends (as the ``$s=x$'' curve
decreases at higher $z$ while the full prediction grows to get closer to it).

\begin{figure}[htb]
\begin{center}
\epsfxsize=8. cm \epsfysize=6. cm {\epsfbox{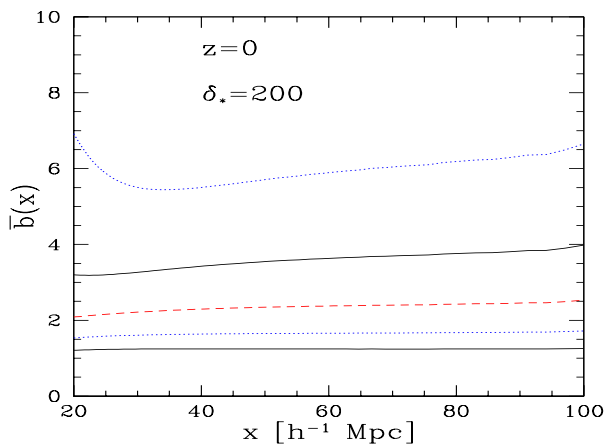}}
\end{center}
\caption{The mean bias $\overline{b}(x)$, as a function of distance $x$,
at redshift $z=0$ for halos defined by the nonlinear density threshold
$\delta_*=200$.
We show our results from Eq.(\ref{bb_Mbin}) for the five mass bins
$10^{13}<M<3\times 10^{13}h^{-1}M_{\odot}$,
$3\times 10^{13}<M<10^{14}h^{-1}M_{\odot}$,
$10^{14}<M<3\times 10^{14}h^{-1}M_{\odot}$,
$3\times 10^{14}<M<10^{15}h^{-1}M_{\odot}$,
and $10^{15}<M<3\times 10^{15}h^{-1}M_{\odot}$,
from bottom to top.}
\label{fig_biasx_Msup_z0}
\end{figure}

In Fig.~\ref{figbiasM_D200} we have considered the bias at the distance
$x=50 h^{-1}$Mpc, which is the typical distance where the large-scale halo
bias is measured in numerical simulations, with box sizes of order
$500 h^{-1}$Mpc \citep{Tinker2010}.
The bias measured in such simulations is almost scale-independent over the
range $20<x<110 h^{-1}$Mpc, as seen for instance in Fig.10 of
\citet{Manera2010} and Fig.17 of \citet{Manera2010a}.
This agrees with the standard large-scale behavior
$b \sim \delta_{L*}/\sigma_q^2$. In our model, there is an additional prefactor
$\sigma_{q,q}(s)/\sigma_{0,0}(x)$, due to halo motions and to the fact that
high wavenumbers do not contribute to halo correlations.
Therefore, we show in Fig.~\ref{fig_biasx_Msup_z0} the dependence on
scale of the bias $\overline{b}(x)$, at $z=0$ for five mass bins $[M_1,M_2]$, defined
as
\beq
\overline{b}(x) = \frac{\int_{M_1}^{M_2} b(M) n(M) \dd M}
{\int_{M_1}^{M_2} n(M) \dd M} .
\label{bb_Mbin}
\eeq
We can check that we obtain a weak dependence on scale over the range
$20<x<100 h^{-1}$Mpc (and we obtain similar results at higher redshift).
At smaller scales, the upturn for the bias of the most massive halos (upper dotted
curve) is due to the fact that the argument of the exponential in Eq.(\ref{b2-def})
becomes large so that the linear approximation is no longer valid.
We shall come back to this point in Sect.~\ref{Impact-of-exponential-nonlinearity}
below. At even smaller scales, of the order of the radius of the halos,
exclusion constraints that we have not taken into account come into play
and imply a halo correlation equal to $-1$. It is clear that our model only applies
to the larger scales, shown in Fig.~\ref{fig_biasx_Msup_z0}, where halos are well
separated.

At larger scales, the bias is difficult to measure (because the two-point
correlation functions are quite small) and actually becomes meaningless.
Indeed, as pointed out in \citet{Valageas2010} for more general initial conditions,
and in previous studies where halos are identified with density peaks
\citep{Coles1989,Lumsden1989,Desjacques2008}, since the halo correlation
is not exactly proportional to the matter correlation (because of the smoothing
at scale $q$ in $\sigma^2_{q,q}(s)$, of halo motions, and of subleading
terms) their zero-crossings do not exactly coincide. This implies that
the ratio $b^2=\xi(M,x)/\xi(x)$ shows wild oscillations and divergent peaks
at the scale where the matter two-point correlation changes sign
(typically at $x \sim 120 h^{-1}$Mpc).
Therefore, at very large scales, $x>100 h^{-1}$Mpc, it is no longer useful to
introduce the ratio $b^2$, and it is best to work with the halo and matter
correlations themselves (note that this conclusion is quite general and not
specific to the model studied here).

\section{Extension to higher density contrasts}
\label{Extension}

\begin{figure*}[htb]
\begin{center}
\epsfxsize=8.5 cm \epsfysize=6 cm {\epsfbox{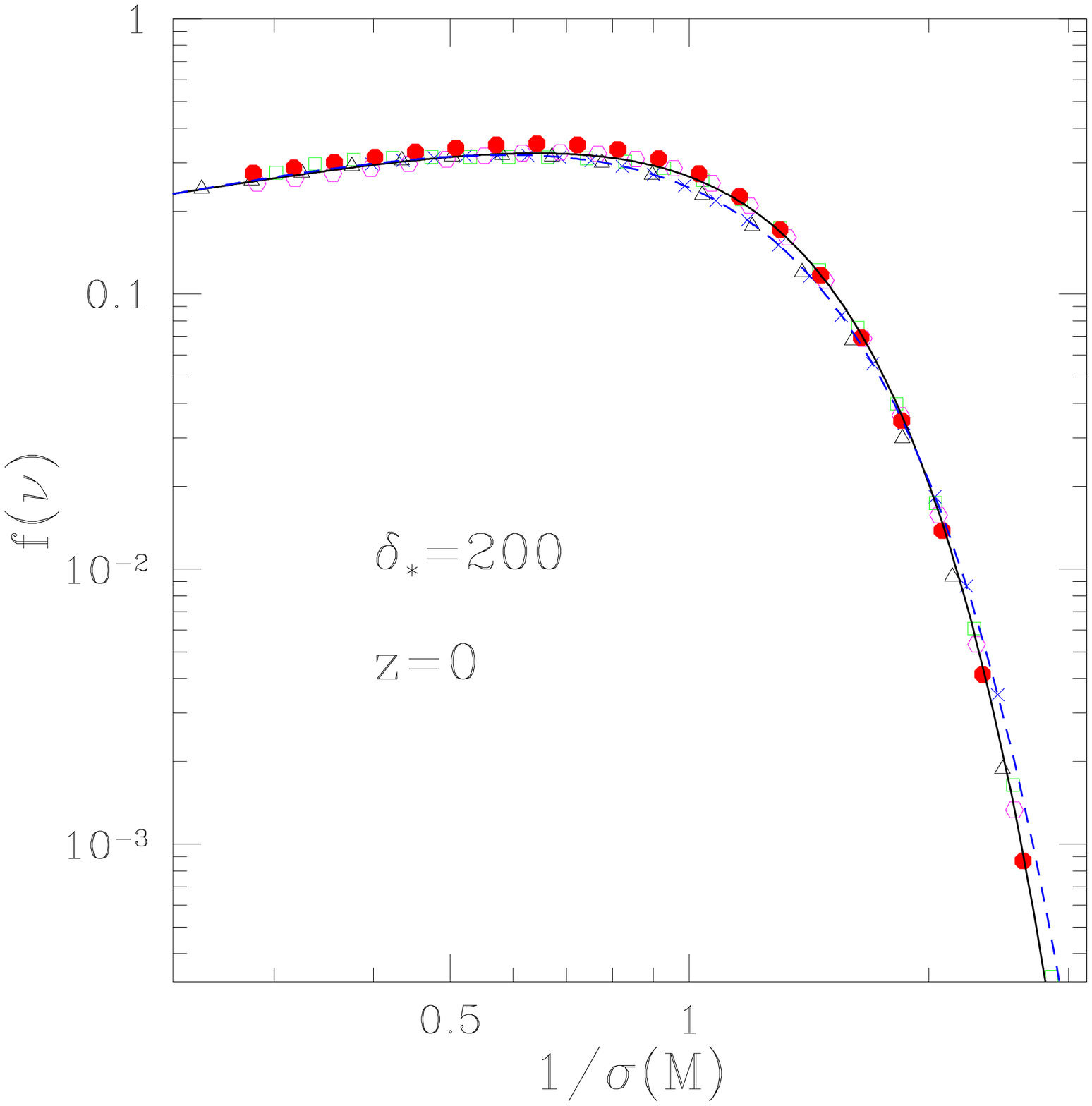}}
\epsfxsize=8.5 cm \epsfysize=6 cm {\epsfbox{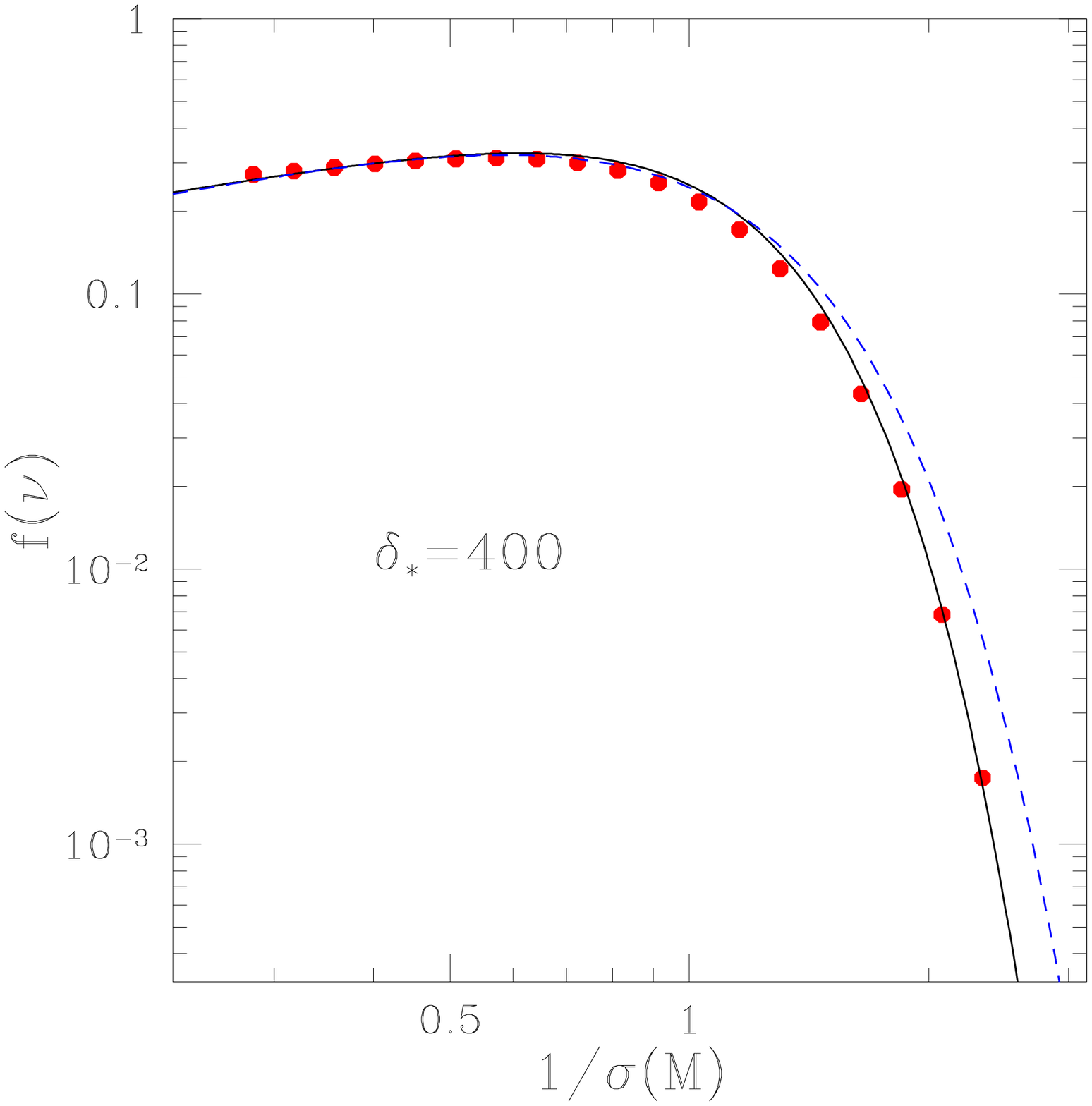}}\\
\epsfxsize=8.5 cm \epsfysize=6 cm {\epsfbox{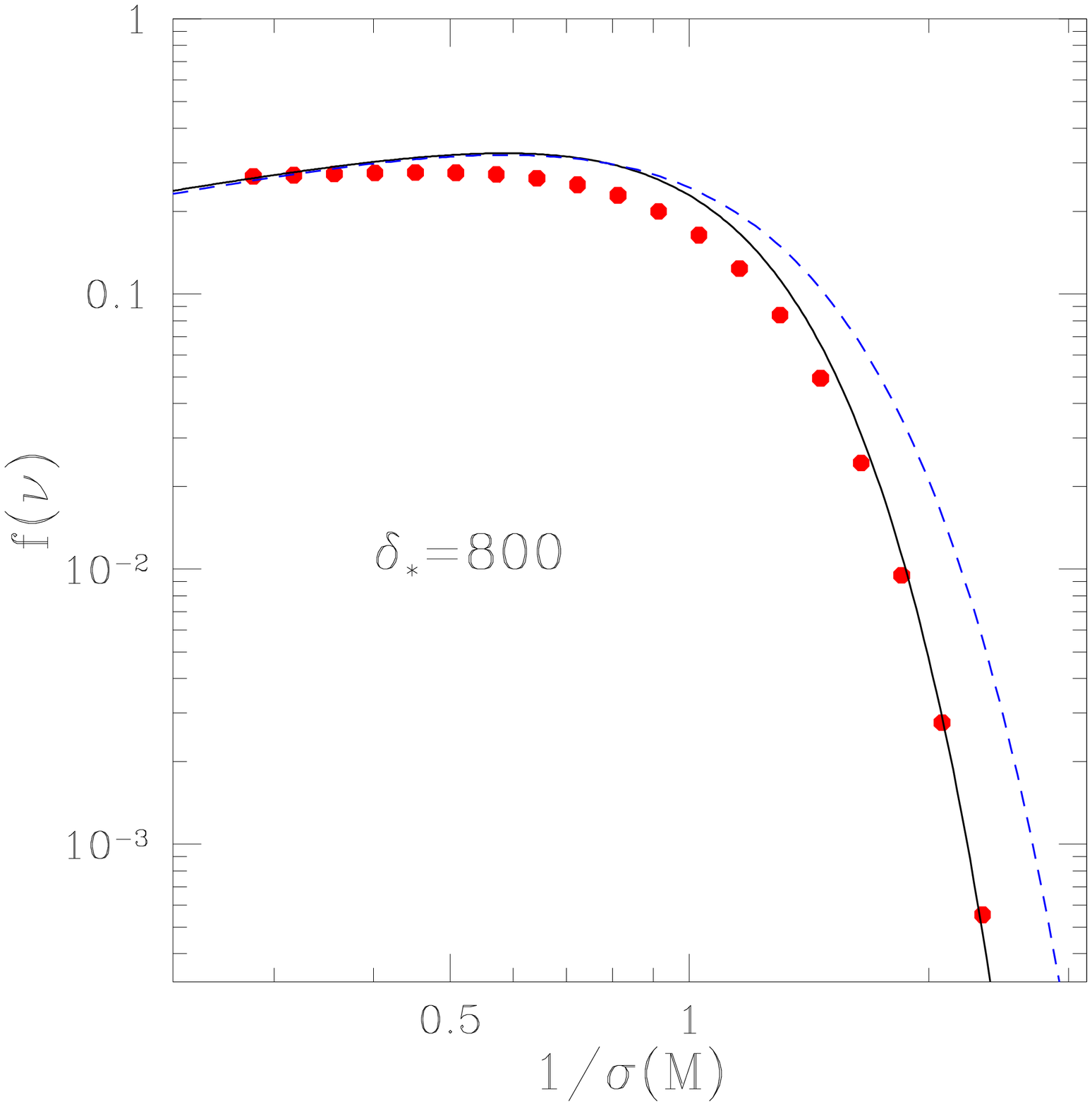}}
\epsfxsize=8.5 cm \epsfysize=6 cm {\epsfbox{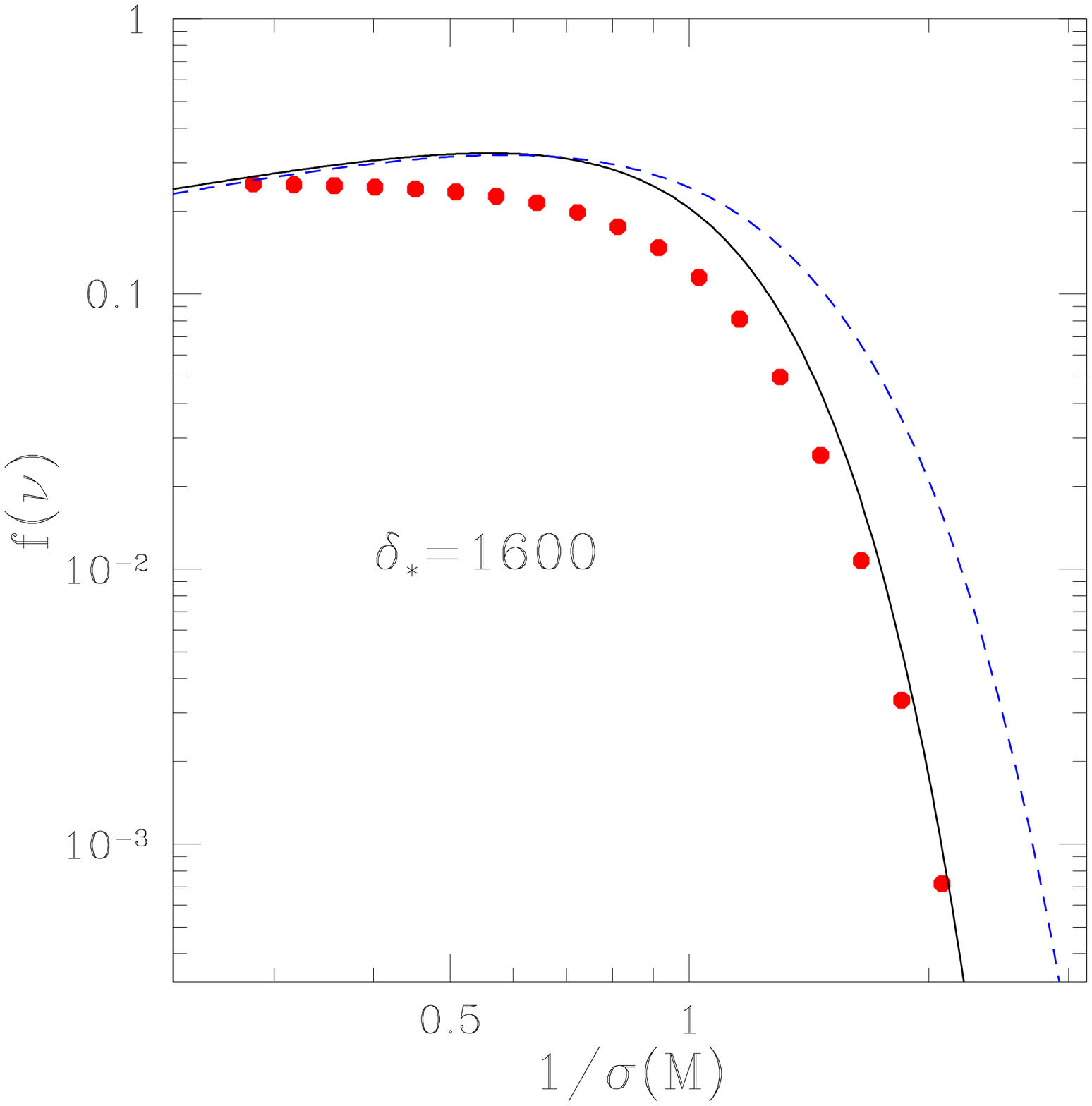}}
\end{center}
\caption{The halo mass functions at redshift $z=0$, for halos defined by a
nonlinear density contrast $\deltas=200, 400, 800$, and $1600$. The solid
line is our model (\ref{fnu-delta-def}), the dashed line is the fit from
\citet{Sheth1999}, which does not change with $\deltas$, and the points are
the results from numerical simulations in \citet{Tinker2008}. The value of
the parameter $\alpha$ in Eq.(\ref{alpha-def}) is set to $\alpha=2.2$.
Here $M=M_{\deltas}$ is the halo mass enclosed within the radius defined
by the density threshold $\deltas$.}
\label{fig_fsig}
\end{figure*}

\begin{figure*}[htb]
\begin{center}
\epsfxsize=6.05 cm \epsfysize=6 cm {\epsfbox{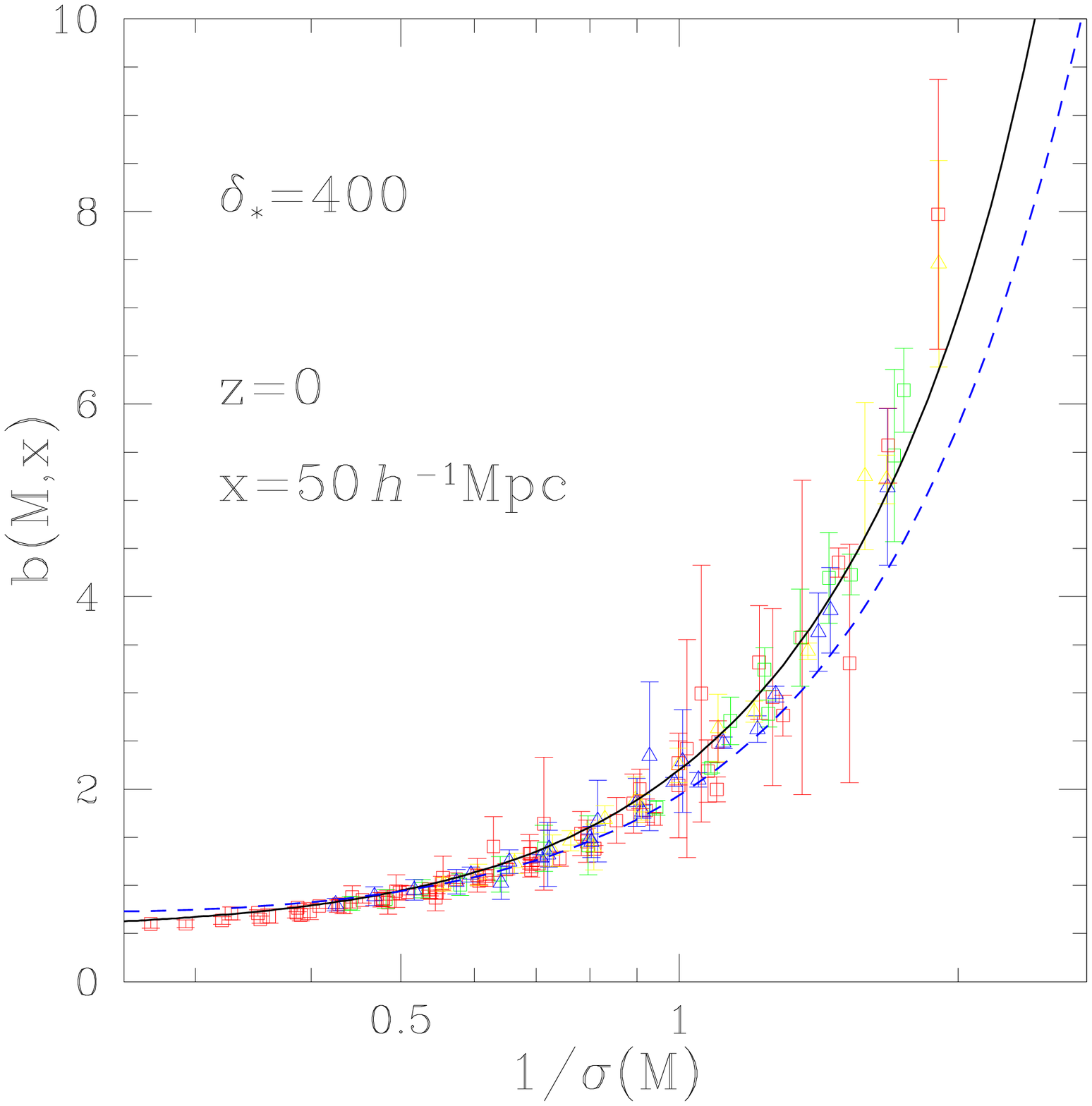}}
\epsfxsize=6.05 cm \epsfysize=6 cm {\epsfbox{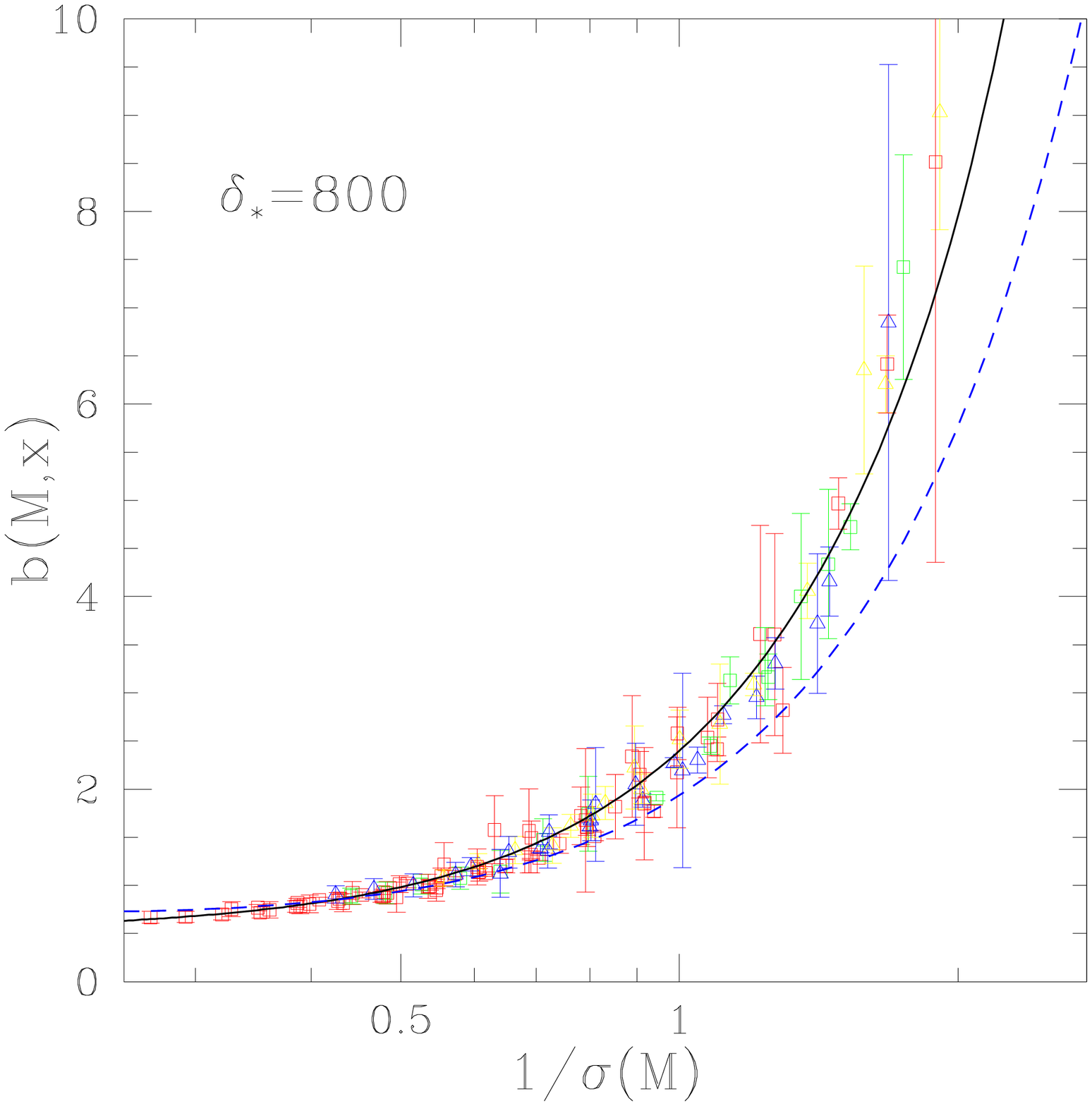}}
\epsfxsize=6.05 cm \epsfysize=6 cm {\epsfbox{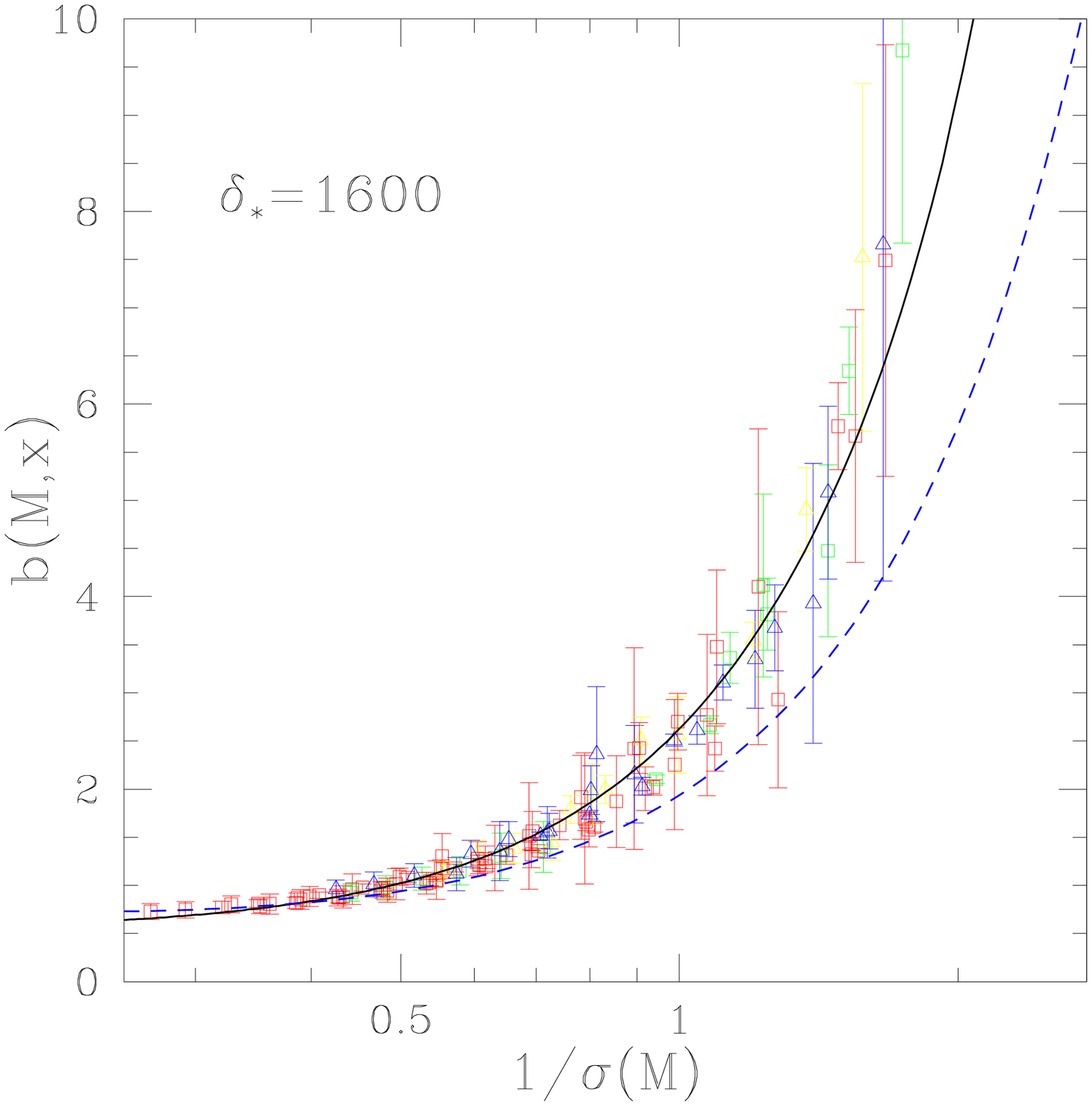}}
\end{center}
\caption{The halo bias $b(M,x)$ as a function of $\sigma(M)$, at distance
$x=50 h^{-1}$Mpc and redshift $z=0$. We show our results for halos defined
by a density threshold $\deltas=400$ (left panel), $\deltas=800$ (middle panel),
and $\deltas=1600$ (right panel). The solid line is our model (\ref{b-delta}),
with $\alpha=2.2$ as for the mass functions of Fig.~\ref{fig_fsig}.
The dashed line is the fit from \citet{SMT2001}, for halos with $\deltas=200$,
which is the same in all panels and in Fig.~\ref{figbiasM_D200}.
The points are the results from numerical simulations of \citet{Tinker2010}.}
\label{figbiasM_D}
\end{figure*}

The results obtained in Sect.~\ref{Bias} applied to halos defined by a
nonlinear density contrast $\deltas$ such that $\deltas \la 200$, so that
the large-mass tails could be derived from the spherical collapse dynamics
associated with Eq.(\ref{qr-F}) (and we focused on the case $\deltas=200$,
which is of practical interest).
As pointed out in \citet{Valageas2009d},
for larger density contrasts shell crossing plays a key role and even in the
rare-event or large-mass limit Eq.(\ref{qr-F}) no longer applies.
Indeed, as discussed in \citet{Valageas2002IV},
because of a strong radial-orbit instability spherical dynamics is no longer
a useful guide. More precisely, the properties of the density field in such
high-density regions are no longer governed by initial configurations that
are spherically symmetric (whence with purely radial motions) and one must
take into account (infinitesimally) small deviations from spherical symmetry,
that are amplified in a non-perturbative manner after collapse.
This is a significant difficulty for precise and robust modeling of halos defined
by high density thresholds. Therefore, we investigate in this paper a very simple
model that tries to bypass this problem by relating inner high-density shells
to outer lower-density radii.

As we have seen in Sect.~\ref{Bias}, in order to obtain a reasonably
successful model for the distribution (i.e., the bias) of dark matter halos,
defined by $\deltas=200$, it is sufficient to derive the large-mass tail,
following the standard ideas of \citet{Kaiser1984} where rare nonlinear objects
are identified in the initial (linear) density field. Then, normalization
conditions constrain the bias of typical halos ($\sigma(M) \ga 1$)
and automatically provide reasonable estimates in this range. The advantage
of this procedure is to bypass a detailed treatment of typical or low-mass
halos, which would require taking into account tidal effects and mergings
(but see \citet{Blanchard1992}).

In this spirit, in order to extend our model to higher density thresholds
$\deltas$, we only need to explicitly consider rare massive halos.
Then, since these halos are isolated and relaxed objects, we assume
that they can be described by a mean mass profile $M(<r)$, with a
mean density within radius $r$ that behaves as
\beq
\rho(<r) = \frac{3 M(<r)}{4\pi\rhob r^3} \propto r^{-\alpha} ,
\;\;\; \mbox{with} \;\;\; 0 < \alpha < 3 ,
\label{alpha-def}
\eeq
where $M(<r)$ is the mass enclosed within radius $r$.
Here we could have used a more detailed density profile, such as the
usual NFW profile \citep{NFW1997}, as described in \citet{Hu2003}.
However, this would also introduce a further
dependence on mass and redshift through the concentration parameter
$c(M,z)$ that parametrizes such profiles. Therefore, in view of the
approximations involved in our approach we think that such refinements are
beyond the scope of this study and we prefer to use a single power-law
index $\alpha$. This should be sufficient as long as we do not consider too
wide a range of contrasts $\deltas$, over which the slope of the density profile
shows significant changes.
Then, from Eq.(\ref{alpha-def}) we can relate the mass $M_{\deltas}$, enclosed
with the radius defined by the nonlinear density contrast $\deltas$,
to $M_{200}$, associated with $\deltas=200$, by
\beq
M_{\deltas} = M_{200} \; \left(\frac{201}{1+\deltas}\right)^{(3-\alpha)/\alpha} .
\label{Mdelta}
\eeq
With the same one-to-one identification, we can define a reduced variable
$\nu_{\deltas}$ as
\beq
\nu_{\deltas}(M_{\deltas}) = \nu_{200}(M_{200}) ,
\label{nu-delta}
\eeq
where $\nu_{200}(M_{200})$ is given by the second Eq.(\ref{fnu-def}) and
$M_{200}$ is obtained as a function of $M_{\deltas}$ through Eq.(\ref{Mdelta}),
and the large-mass tail of the halo mass function is still given by
$n_{\deltas}(M_{\deltas}) \sim e^{-\nu_{\deltas}^2/2} = e^{-\nu_{200}^2/2}$.
This one-to-one identification of rare massive halos also implies that their
spatial distribution remains the same, independently of the choice of $\deltas$,
so that the large-mass two-point correlation and bias are still given by
Eqs.(\ref{xi-M}) and (\ref{b2-def}),
\beq
b^2_{{\rm r.e.};\deltas}(M_{\deltas},x) = b^2_{{\rm r.e.};200}(M_{200},x) .
\label{b-re-delta}
\eeq

Next, in order to describe typical halos we again take advantage of normalization
constraints. First, to ensure that the mass function remains normalized to unity
as in Eq.(\ref{fnu-norm}), we choose the approximation
\beq
n_{\deltas}(M_{\deltas})\dd M_{\deltas} = \frac{\rhob}{M_{\deltas}} \, f(\nu) \,
\frac{\dd\nu}{\nu} ,
\label{fnu-delta-def}
\eeq
where the reduced function $f(\nu)$ is still given by Eq.(\ref{fnu-fit}) and
the reduced variable $\nu$ is given by Eq.(\ref{nu-delta}). Here the mass
$M_{200}$ is still defined by Eq.(\ref{Mdelta}) but since there is no longer
a one-to-one identification (except at very large mass, as explained below)
$M_{200}$ must be seen as an ``effective mass'' rather than the mass
within $\deltas=200$ of the same individual object.
This prescription automatically
provides both the right large-mass cutoff, as explained below Eq.(\ref{nu-delta})
(provided the mean profile (\ref{alpha-def}) is correct) and the normalization
(\ref{fnu-norm}). Note that this is a normalization in terms of mass, which ensures
that by counting such halos we recover the mean matter density of the Universe.
Thus, a fixed range $[\nu_1,\nu_2]$ is associated with the same fraction of
matter, independently of the choice of $\deltas$, but it is divided into a larger
number of smaller objects for larger $\deltas$, because of the prefactor
$\rhob/M_{\deltas}$ in Eq.(\ref{fnu-delta-def}).
Indeed, a given $\nu$ is associated to a fixed $M_{200}$, through the second
Eq.(\ref{fnu-def}), and to a mass
$M_{\deltas} \propto (1+\deltas)^{-(3-\alpha)/\alpha}$ through Eq.(\ref{Mdelta}).
Thus, through the normalization constraint and the form (\ref{fnu-delta-def})
we include some dependence of the halo multiplicity on the choice of $\deltas$
(as expected, a higher threshold $\deltas$ leads to a larger number of objects),
and the one-to-one identification used in Eqs.(\ref{Mdelta})-(\ref{b-re-delta})
only applies in an asymptotic sense at high mass.

Second, to ensure that the halo bias also remains normalized to unity,
as in Eq.(\ref{b-fnu}), we also add to the large-mass term
$b_{{\rm r.e.};\deltas}(M_{\deltas},x)$ a mass-independent term $b_{0;\deltas}(x)$,
as in Eq.(\ref{bM-def}). Then, thanks to the choice (\ref{fnu-delta-def}) we
find that the normalization (\ref{b-fnu}) yields
$b_{0;\deltas}(x)=b_{0;200}(x)$. Combining with Eq.(\ref{b-re-delta}) this gives
for the total halo bias,
\beq
b_{\deltas}(M_{\deltas},x) = b_{200}(M_{200},x) .
\label{b-delta}
\eeq
This automatically satisfies both the expected large-mass behavior
(where the halos labeled by $\deltas$ and $200$ are the same objects)
and the normalization constraint. Since at moderate and low mass we no
longer have a one-to-one identification between halos, as we have seen from the
mass function (\ref{fnu-delta-def}), the equality (\ref{b-delta}) should {\it not} be
seen as the result of a strict one-to-one identification, even though it happens
to give a similar expression.

We first show in Fig.~\ref{fig_fsig} the halo mass functions we obtain with our
simple model at redshift $z=0$.
In the following, for simplicity we do not write the subscript $\deltas$,
so that in each panel of Fig.~\ref{fig_fsig} the mass $M$ is actually the
mass $M_{\deltas}$ enclosed within the radius defined by the density threshold
$\deltas$, as labeled in each plot.

For $\deltas>200$ we set the parameter $\alpha$ of Eq.(\ref{alpha-def}) to
$\alpha=2.2$. 
Within our approach, this is the only new parameter that is needed to obtain the
mass functions for arbitrary $\deltas$ from the one at $\deltas=200$ given by
Eq.(\ref{fnu-fit}). We choose a value such that the large-mass tail is correctly
reproduced for $200\leq\deltas\leq 1600$. 
This gives a reasonable value for the slope of the density profile around the
virial radius (i.e. $\rho(<r) \sim r^{-2.2}$).
For larger values of the density threshold, which correspond to smaller radii
within massive halos, a smaller value of $\alpha$ would probably be required
in order to follow the flattening of the halo profile, down to $\alpha \simeq -1$.
Since for practical purposes one usually considers density thresholds in the
range $170\leq\deltas\leq 500$ to define dark matter halos, we do not investigate
further this point.
As seen in Fig.~\ref{fig_fsig}, our simple model is already sufficient to reproduce
the shift of the large-mass tail over the range $200\leq \deltas \leq 1600$.
This is most clearly seen by comparison with the fixed dashed line that applies
to $\deltas=200$, taken from \citet{Sheth1999}.

For high values of the threshold $\deltas$, especially in the lower right panel
at $\deltas=1600$, it appears that our model overestimates the mass function
for intermediate-mass halos, $\sigma(M) \sim 1.5$. On the other hand, it
seems that in numerical simulations some fraction of matter is lost
as we consider higher thresholds $\deltas$, so that the normalization
(\ref{fnu-norm}) no longer holds (the integral would be smaller than unity
for large $\deltas$).
This may have a physical meaning, for instance if some regions of space
are smooth and show a maximum density contrast, in which case they are
no longer included as we select a higher threshold. A second factor is
that at some level the distribution of the matter content of the Universe 
over a set of halos is not a very well defined procedure. 
More precisely, even though one may always build a well-defined algorithm
to redistribute particles in a collection of halos within numerical simulations,
to some degree it involves some arbitrariness from a physical point of view.
This is clear from the fact that one cannot build a partition of a 3D box
with spheres, and this means that the description of the matter distribution
in terms of spherical halos can only be approximate.
This may also lead to a violation of the normalization (\ref{fnu-norm}).
A detailed investigation of such effects is beyond the scope of this paper
and we prefer to stick to our simple model and to fixed normalizations such 
as Eq.(\ref{fnu-norm}).
For practical purposes, the accuracy of the model (\ref{fnu-delta-def}) should
be sufficient, and appears to be quite satisfactory in view of its simplicity.

We compare in Fig.~\ref{figbiasM_D} the halo bias predicted by our model,
Eq.(\ref{b-delta}), with results from numerical simulations \citep{Tinker2010},
for the density thresholds $\deltas=400, 800$, and $1600$.
Although the theoretical curve corresponds to $z=0$, in order to increase
the statistics we plot the numerical data obtained for $0\leq z \leq 2.5$,
since the dependence on redshift is quite weak (see Fig.~\ref{figbiasM_D200}).
Of course, for the parameter $\alpha$ we use the same value as the
one used for the mass functions shown in Fig.~\ref{fig_fsig}, $\alpha=2.2$.

We can see that as the density threshold $\deltas$ increases the bias
at fixed $\sigma(M)$ grows, especially at large masses. This is most clearly
seen through the comparison with the fixed dashed line, which corresponds to
the fit from \citet{SMT2001} for halos defined by $\deltas=200$.
This moderate
dependence on $\deltas$ is also reproduced by our simple model (\ref{b-delta}),
which shows a reasonable agreement with N-body simulations over this range,
$200\leq \deltas \leq 1600$. In fact, for a constant parameter $\alpha$,
as in Fig.~\ref{figbiasM_D}, we can see from Eq.(\ref{b-delta}) that as
$\deltas$ grows the curve $b(M)$ is simply shifted by a uniform translation
towards the left in the $(\ln M,b)$ plane, since at fixed bias $b$, whence
at fixed ``effective mass'' $M_{200}$, we have
$\ln M = \ln M_{200} +  \frac{3-\alpha}{\alpha} \ln [201/(1+\deltas)]$.
Since $\sigma(M)$ is not exactly a power law, this horizontal translation
is not exactly uniform in the $(\ln(1/\sigma),b)$ plane of Fig.~\ref{figbiasM_D}.
As noticed above, in order to cover a larger range of density thresholds
$\deltas$, or to improve the accuracy, it would be necessary to use a slope
$\alpha$ in Eq.(\ref{alpha-def}) that runs with $M$ and $\deltas$, in order
to take into account the dependence on mass of the halo profile and the
flattening of the slope at inner radii.
However, we can see in Fig.~\ref{figbiasM_D} that the simple model with
a constant value, $\alpha=2.2$, already provides a good match to numerical
simulations.

An advantage of this simple approximation is that it is straightforward to
satisfy the normalization (\ref{b-fnu}), as shown by the simple consequence
(\ref{b-delta}). This ensures that our underlying halo model is self-consistent,
in the sense that if we describe the matter distribution through a standard
halo model \citep{Cooray2002}, but with arbitrary threshold $\deltas$ to
define the halos, by integrating over these halos we recover both the
total matter density (through the halo mass function) and the matter
two-point correlation (through the halo bias).
This means that, despite the approximate nature of such halo descriptions
noticed above in the discussion of Fig.~\ref{fig_fsig}, our model automatically
satisfies these two integral constraints.

We think such properties should be taken into account in the building of models
for the matter distribution. As seen in this work, they can serve as a useful guide,
which can be sufficient to provide reasonable quantitative estimates over
tightly constrained domains. A second benefit is that they avoid introducing
small inconsistencies, which may lead to spurious quantitative discrepancies
for quantities that would be computed in later steps by integrating over
the halo populations.

\section{Impact of halo motions}
\label{Impact-of-halo-motions}

\begin{figure}[htb]
\begin{center}
\epsfxsize=8.5 cm \epsfysize=6.5 cm {\epsfbox{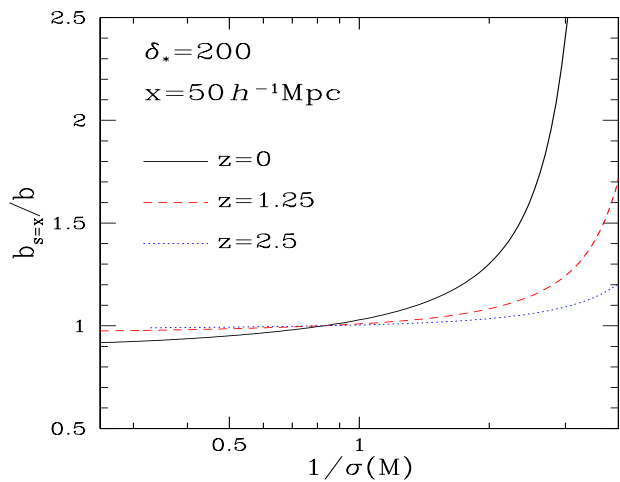}}
\end{center}
\caption{The ratio $b_{s=x}/b$, as a function of $\sigma(M)$, at redshifts
$z=0, 1.25$, and $2.5$. The bias $b$ is the one obtained in Sect.~\ref{Bias} 
and shown in Fig.~\ref{figbiasM_D200}, whereas $b_{s=x}$ is obtained by
making the approximation $s=x$ in Eq.(\ref{b2-def}).}
\label{fig_rbiasM_sx}
\end{figure}

We now take advantage of the simple analytical model presented in
Sect.~\ref{Bias} to estimate the impact of halo motions on their observed
two-point correlation. Thus, focusing on halos defined by the nonlinear density
threshold $\deltas=200$, we plot in Fig.~\ref{fig_rbiasM_sx} the ratio
$b_{s=x}/b$, where $b$ is the bias obtained in Sect.~\ref{Bias} and
shown by the solid line in Fig.~\ref{figbiasM_D200}, while $b_{s=x}$ is the bias
obtained with the approximation $s=x$ (i.e. we substitute $s\rightarrow x$ in
Eq.(\ref{b2-def})), which was plotted as the red dashed line in
Fig.~\ref{figbiasM_D200}.
Therefore, this ratio measures the effect on the bias of halo motions, which are
neglected in the usual approach \citep{Kaiser1984} or in the
peak-background split method \citep{Cole1989,MW1996}.

For rare massive halos, where the asymptotic expression (\ref{b2-def}) applies,
we have $s>x$ within our complete approach, as can be checked from
Eq.(\ref{s-x-M}). Thus, as could be expected, rare massive halos tend to move
closer because of their mutual gravitational attraction. Although this might seem
contradictory with large-scale homogeneity (one may ask how could all halos
move closer to each other ?), this is not the case.
For instance, let us consider an infinite regular 3D grid of cell size
$L$, and let us put two halos separated by a small distance, $\ell \ll L$, around
each vertex (so that the vertex is the center of the pair). Then, if at a later
time we independently shrink each pair around its vertex,
$\ell \rightarrow \ell'$ with $\ell'<\ell$,
we can see that on the mean halos have moved closer. Indeed, the distances
between different pairs have not changed while they have been reduced within
each pair. However, the system has clearly remained homogeneous on large scales
(there is not coherent shrinking of all halos positions towards a central point, as
one might have been afraid of). 
This simple picture shows how small scale motions can have a nonzero effect
on the mean, while preserving large-scale statistical homogeneity.

\begin{figure}[htb]
\begin{center}
\epsfxsize=8.5 cm \epsfysize=6.5 cm {\epsfbox{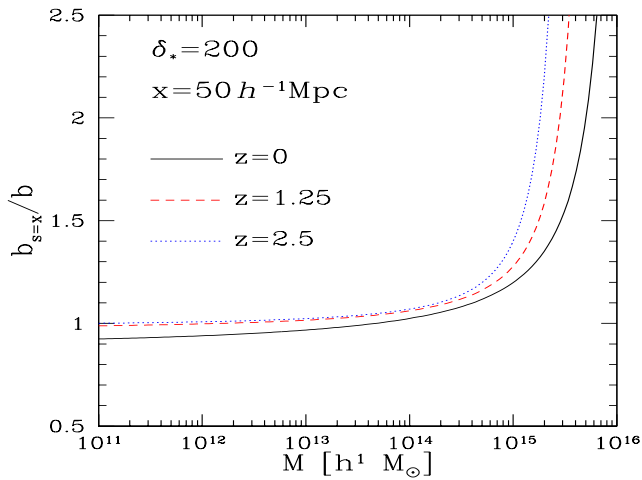}}
\end{center}
\caption{The ratio $b_{s=x}/b$, as in Fig.~\ref{fig_rbiasM_sx}, but as a function
of $M$, at redshifts $z=0, 1.25$, and $2.5$.}
\label{fig_rbias_M_sx}
\end{figure}

Then, it is clear from Eq.(\ref{b2-def}) that the predicted bias, or the predicted
two-point correlation function, of rare massive halos is smaller when we take into
account the fact that $s>x$ than when we neglect this distinction.
Indeed, cross-correlations, such as $\sigma^2_{q,q}(s)$, of the linear density
fluctuations at scale $q$ over a large distance $s$ are decreasing functions of $s$.
This implies that by making the approximation ``$s=x$'' we overestimate
the initial cross-correlation between massive halos observed at distance $x$,
and this explains why the ratio $b_{s=x}/b$ is larger than unity at large masses
in Fig.~\ref{fig_rbiasM_sx}. This effect increases at larger masses, which have
a large bias and show a stronger sensitivity to the large-scale correlations.
Of course, this agrees with the behavior found in Fig.~\ref{figbiasM_D200}.
At small masses the ratio $b_{s=x}/b$ becomes slightly smaller than unity.
This is a direct consequence of the normalization constraint (\ref{b-fnu}), which
implies that the greater value of the bias at high mass, within the
approximation ``$s=x$'', must be compensated by a smaller value at small mass.
The effect is rather small in this range since massive halos are rare so that their
bias does not contribute much to the normalization (\ref{b-fnu}).

We can see in Fig.~\ref{fig_rbiasM_sx} that the effect of halo motions is smaller at
higher redshift for a fixed value of $\sigma(M)$, which actually corresponds to
a smaller mass at higher $z$. Thus, if one only considers typical halos, or a
population defined by a fixed upper bound on $1/\sigma(M)$ (this roughly
corresponds to a fixed comoving number density), the average halo motion can
be neglected at high $z$. We show in Fig.~\ref{fig_rbias_M_sx} the same
ratio $b_{s=x}/b$ at redshifts $z=0, 1.25$, and $2.5$, but as a function of
$M$ instead of $1/\sigma(M)$. Then, we can see that the effect of the average
halo motion now increases at higher redshift, for a fixed mass $M$.

At $z=0$, neglecting this effect leads to an overestimation of the bias $b(M)$
by a factor $1.5$ at $M \sim 3 \times 10^{15} h^{-1}M_{\odot}$, that is,
$\sigma(M) \sim 0.34$. Therefore, it cannot be neglected for massive halos.

\section{Impact of exponential nonlinearity}
\label{Impact-of-exponential-nonlinearity}

\begin{figure}[htb]
\begin{center}
\epsfxsize=8.5 cm \epsfysize=6.5 cm {\epsfbox{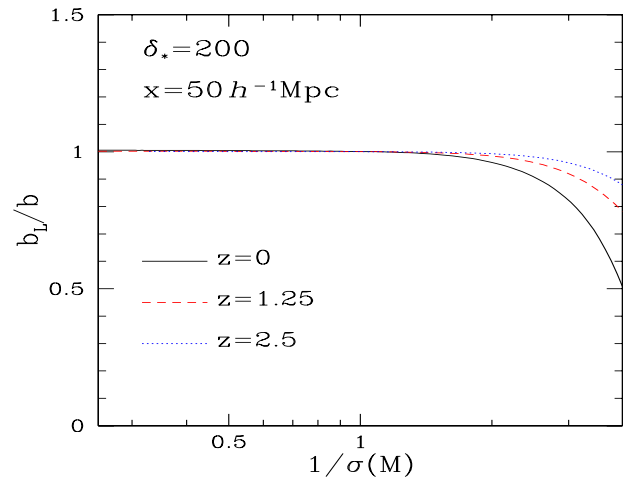}}
\end{center}
\caption{The ratio $b_L/b$, as a function of $\sigma(M)$, at redshifts
$z=0, 1.25$, and $2.5$. The bias $b$ is the one obtained in Sect.~\ref{Bias} 
and shown in Fig.~\ref{figbiasM_D200}, whereas $b_L$ is obtained by
using the ``linear'' approximation (\ref{b2L-def}) in Eq.(\ref{bM-def}).}
\label{fig_rbiasM_lin}
\end{figure}

\begin{figure}[htb]
\begin{center}
\epsfxsize=8.5 cm \epsfysize=6.5 cm {\epsfbox{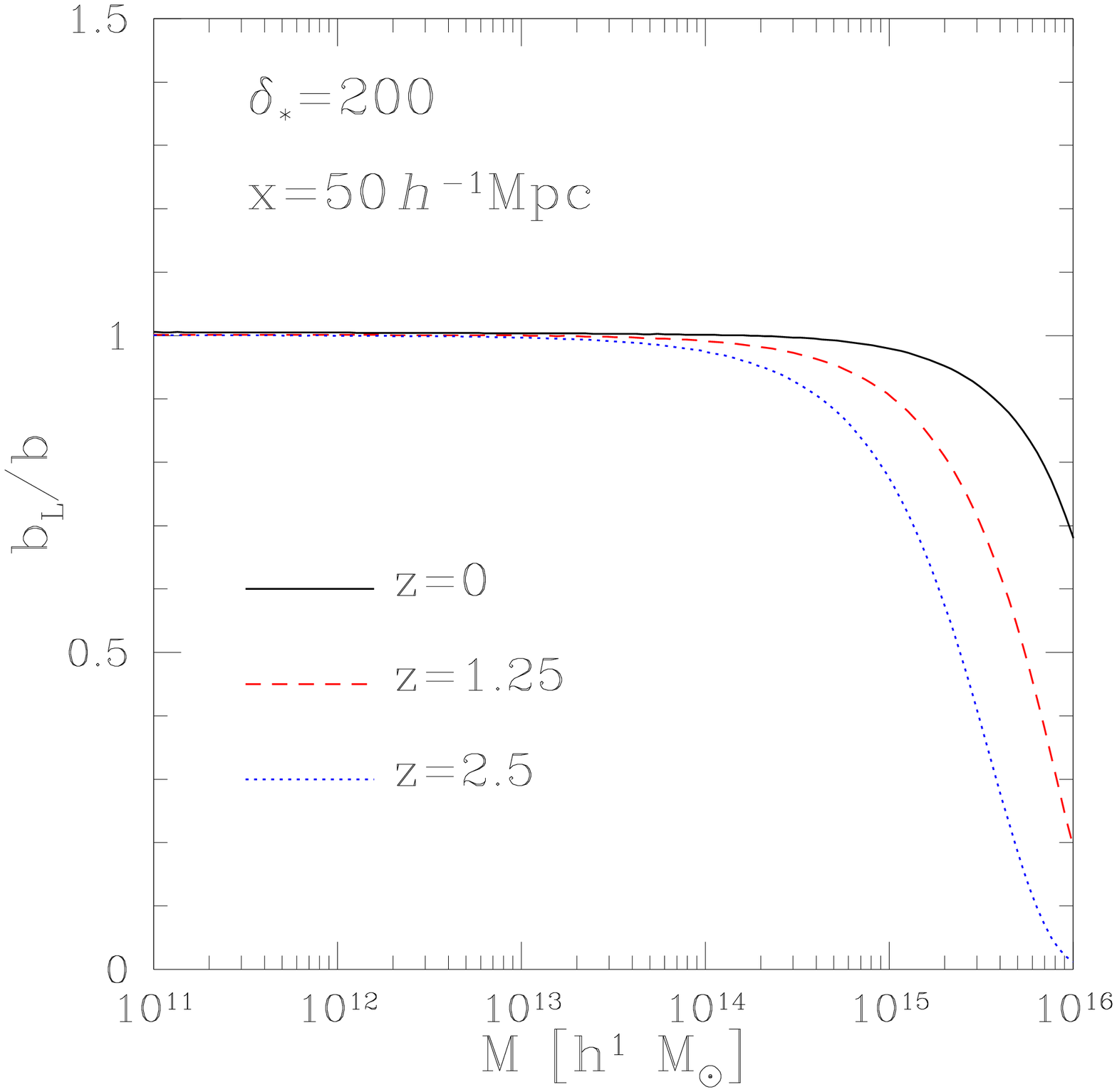}}
\end{center}
\caption{The ratio $b_L/b$, as in Fig.~\ref{fig_rbiasM_lin}, but as a function
of $M$, at redshifts $z=0, 1.25$, and $2.5$.}
\label{fig_rbias_M_lin}
\end{figure}

At large distance cross-correlations such as $\sigma^2_{q,q,s}$ become very
small and one can expand the exponential in Eq.(\ref{b2-def}), to obtain
the linearized form
\beq
b^2_{\rm r.e.;L}(M,x) = \frac{\delta_{L*}^2}
{\sigma^2_q(\sigma^2_q+\sigma^2_{q,q}(s))} \, \frac{\sigma^2_{q,q}(s)}
{\sigma^2_{0,0}(x)} .
\label{b2L-def}
\eeq
At zeroth-order over $\sigma^2_{q,q}(s)$, where $s=x$, this gives the
well-known result of
\citet{Kaiser1984}, $b_{\rm r.e.;L}(M,x) \sim \delta_{L*}/\sigma_q^2$,
which shows how the bias of massive halos grows as $\propto 1/\sigma_q^2$,
in agreement with Fig.~\ref{figbiasM_D200}. 
Linearized expressions such as (\ref{b2L-def}), that is, where one looks for an
expression of the halo two-point correlation function, or of the halo power
spectrum, that is linear over the matter two-point correlation or power
spectrum, are widely used. In terms of the ``local bias model''
\citep{Fry1993,Mo1997,Manera2010a}, this
also corresponds to the approximation of ``linear bias'', where the halo
density field, $n_M(\vx)$, is written as a linear function of the matter density
field, such as $n_M(\vx) = \nb_M  [1 + b_1(M) \delta_M(\vx) ]$
(where $\delta_M(\vx)$ is the matter density contrast  smoothed on scale
$M$), with a constant linear bias factor $b_1$.
However, as pointed out by \citet{Politzer1984}, even when the cross-correlation
$\sigma^2_{q,q}(s)$ is small the argument of the exponential (\ref{b2-def}) can
be large, because of the factor $\delta_{L*}^2/\sigma_q^4$. For instance, at
fixed scales $q$ and $s$ the argument grows as $1/\sigma^2$ as the amplitude
of the linear matter power spectrum decreases. 
Therefore, we investigate in this section the impact of the exponential
non-linearity (\ref{b2-def}) on the halo bias.

Thus, we show in Fig.~\ref{fig_rbiasM_lin} the ratio $b_L/b$, where $b$ is
again the bias obtained in Sect.~\ref{Bias} and Fig.~\ref{figbiasM_D200},
while $b_L$ is the bias obtained with the approximation (\ref{b2L-def}),
which we substitute into Eq.(\ref{bM-def}). We still take into account halo
motions through Eq.(\ref{s-x-M}), as well as normalizations to unity.
Therefore, this ratio measures the effect of the exponential non-linearity
(\ref{b2-def}).
In agreement with the previous discussion, we can see in Fig.~\ref{fig_rbiasM_lin}
that by linearizing the expression (\ref{b2-def}) we underestimate the
halo bias at large masses. Indeed, the argument in Eq.(\ref{b2-def}) scales as
$\sim \delta_{L*}^2\sigma^2_{q,q,s}/\sigma_q^4 \propto 1/\sigma_q^2$,
which grows at large mass.
However, this effect is only significant at very large masses and typical
objects are not affected by this linearization.
As shown by Figs.~\ref{fig_rbiasM_lin} and \ref{fig_rbias_M_lin}, this effect
decreases at higher redshift for a fixed value of $\sigma(M)$, but increases
at higher redshift for a fixed mass $M$.
Therefore, although for typical halos such a ``linearization'' of the bias
is a very good approximation, for very massive objects it can lead to a
significant underestimation, especially at high redshift.
As noticed in Fig.~\ref{fig_biasx_Msup_z0} (upper dotted curve), this effect also
becomes more important at smaller distance $x$, as $\sigma^2_{q,q,s}$ is
larger.

\section{Conclusion}
\label{Conclusion}

We have described in this paper a simple analytical model for the
bias of dark matter halos. It extends previous works, which focused on
halos defined by their virial density contrast ($\deltas \sim 200$), to the
more general case of halos defined by an arbitrary density threshold in
the range $200\leq\deltas\leq 1600$. This is coupled to a model for the
halo mass function that is also generalized to these density thresholds,
and we have checked that these simple models yield a good agreement
with numerical simulations at redshifts $0\leq z \leq 2.5$.

Our approach also improves over some previous works by including
the effect of halo motions on their two-point correlation function.
This arises from the fact that massive halos move closer on the mean, because
of their mutual gravitational attraction, which implies that their mean distance 
at a given time is smaller than the Lagrangian separation of the two regions 
they originate from in the primordial density field.
We estimate this effect from the linear displacement of halos, which provides
a simple approximation that could be easily used in more complex cases, such
as non-Gaussian initial conditions.

Another feature of our model is that it contains no free parameters for the bias.
More precisely, the only parameters that appear are related to the halo mass
functions. As in most other approaches, a few parameters are needed to
describe the shape of the mass function (of halos defined by $\deltas=200$)
at low and intermediate masses, but the large mass tail is governed by the
cutoff $e^{-\nu^2/2}$ without further tuning. Then, the extension to halo
populations defined by larger density contrasts only involves a single
parameter, $\alpha\simeq 2.2$, which is related to the slope of halo density
profiles around the virial radius. Then, no further parameters are required to
compute the bias of these halo.

As stressed in this paper, this is possible thanks to the use of integral constraints
for both the halo mass function and bias. This allows us to focus on rare and
massive objects, where reliable predictions can be derived. Then, the behavior
of intermediate and small halos, which is beyond the reach of analytical
approaches because of strong tidal effects and mergings, is strongly constrained
by these integral conditions, and we have shown that this is sufficient to
build simple and efficient models.
Of course, this simplicity has a cost: we cannot expect a priori high accuracy at
very low mass, and there is no systematic procedure (such as expansions over
some parameters) to improve the model up to arbitrarily high accuracy.
However, we have seen that we already obtain a good agreement with numerical
simulations, and this approach should be quite robust.
Another advantage is that by explicitly taking into account such integral
constraints we make sure our model is self-consistent, and we avoid any
risk to introduce spurious discrepancies that may arise in integral quantities
because of small inconsistencies.

Finally, we have evaluated the quantitative impact on halo bias of two
common approximations, i) neglecting halo motions, and ii) linearizing the
halo two-point correlation over the matter power spectrum. This could be
useful to check the range of validity of these approximations.

It would be interesting to generalize this work to more complex cases, such as
non-Gaussian initial conditions. However, we leave this to future studies.

\acknowledgements

The author thanks J. L. Tinker for sending his numerical results for the
halo bias, used in Figs.~\ref{figbiasM_D200} and \ref{figbiasM_D}.

\bibliographystyle{aa} % style aa.bst
\bibliography{15699}

\end{document}